\documentclass[vecphys]{svmult}

\usepackage{makeidx}         
\usepackage{graphicx}        
\usepackage{multicol}         
\usepackage[bottom]{footmisc}
\usepackage{epstopdf}

\makeindex             

\begin{document}

\title*{X-ray emission from black-hole binaries}
\author{Marat Gilfanov\inst{1,}\inst{2}}
\institute{Max-Planck-Institute for Astrophysics, Garching, Germany 
and\\
Space Research Institute, Moscow, Russia\\
\texttt{gilfanov@mpa-garching.mpg.de}}

\maketitle

\abstract{
The properties of X-ray emission from accreting black holes are reviewed.  The contemporary observational picture and current status of theoretical understanding of accretion and  formation of X-ray radiation in the vicinity of the compact object are equally in the focus of this chapter. The emphasis is made primarily on common properties and trends rather than on peculiarities  of individual objects and details of particular theoretical models. The chapter starts with discussion of the geometry of the accretion flow, spectral components in X-ray emission and black hole spectral states. The prospects and diagnostic potential of X-ray polarimetry are emphasized.  Significant attention is paid to the discussion of variability of X-ray emission in general and of different spectral components -- emission of the accretion disk, Comptonized radiation and reflected component. Correlations between spectral and timing characteristics of X-ray emission are reviewed and discussed in the context of theoretical models. 
Finally, a comparison with accreting neutron stars is made.
}

\section{Introduction}
\label{sec:intro}

The gravitational energy of matter dissipated in the accretion flow around 
a compact object of stellar mass is primarily converted to photons of  X-ray
wavelengths. 
The lower limit on the characteristic temperature of the spectral energy distribution of the emerging  radiation can be estimated assuming the most
radiatively efficient configuration -- optically thick accretion
flow. Taking into account that the size of the emitting region is
$r\sim 10r_g$ ($r_g$ is the gravitational radius) and assuming a black body
emission spectrum one obtains: 
$kT_{bb}\approx \left(L_X/\sigma_{SB}\pi r^2\right)^{1/4}\approx 
1.4\, L_{38}^{1/4}/M_{10}^{1/2}$ keV. 
It is interesting (and well known) that $T_{bb}$ scales as
$\propto M^{-1/2}$. This is confirmed very well by the measurements of the
disk emission temperature in stellar mass systems and around
supermassive  black holes in AGN. It is also illustrated, albeit less dramatically,  by
the comparison of the soft state spectra of black holes and neutron
stars, as discussed later in this chapter.
The upper end of the relevant temperature range is achieved in the
limit of the optically thin emission.
It is not unreasonable to link it to the virial temperature\index{virial temperature} of
particles near the black hole, 
$kT_{vir}=GMm/r\propto mc^2/(r/r_g)$.
Unlike the black body temperature this quantity  does
not depend on the mass of the compact object, but does depend on the mass
of the particle $m$. For electrons $T_{vir}\sim 25(r/10r_g)^{-1}$
keV and it is correspondingly $m_p/m_e=1836$
times higher for protons. Protons and ions are the main energy reservoir in the accretion flow, but for all plausible mechanisms of spectral formation it is the temperature of electrons that determines the spectral energy distribution of the emerging radiation. The latter  depends on the poorly constrained efficiency of the energy exchange between electrons and protons in the plasma near the compact object.
The values of the electron temperature typically derived from the spectral fits to the hard
spectral component in accreting black holes,  $kT_e\sim 50-150$ keV, are comfortably within the range defined by the two virial temperatures. However, the concrete value of $kT_e$ and its universality in a diverse sample of objects and broad range of luminosity levels still remains unexplained from first principles, similar to those used in  the derivation of the temperature of the optically thick soft component.  

Broadly speaking, significant part of, if not the entire diversity of the spectral behavior observed in accreting black holes and neutron stars can be explained by the changes in the proportions in which the gravitational   energy of the accreting matter is dissipated in the optically thick and optically thin parts of the accretion flow.   The particular mechanism driving these changes is however unknown -- despite significant progress in MHD simulations of the accretion disk achieved in recent years (see Chapter 10)  there is no acceptable global model of accretion onto a compact object.
To finish this introductory note, I will mention that non-thermal processes in optically thin media (e.g. Comptonization on the non-thermal tail of the electron distribution)
may also contribute to the X-ray emission from black holes in some spectral states.

\begin{figure}
\centering
\hbox{
\includegraphics[width=0.48\textwidth]{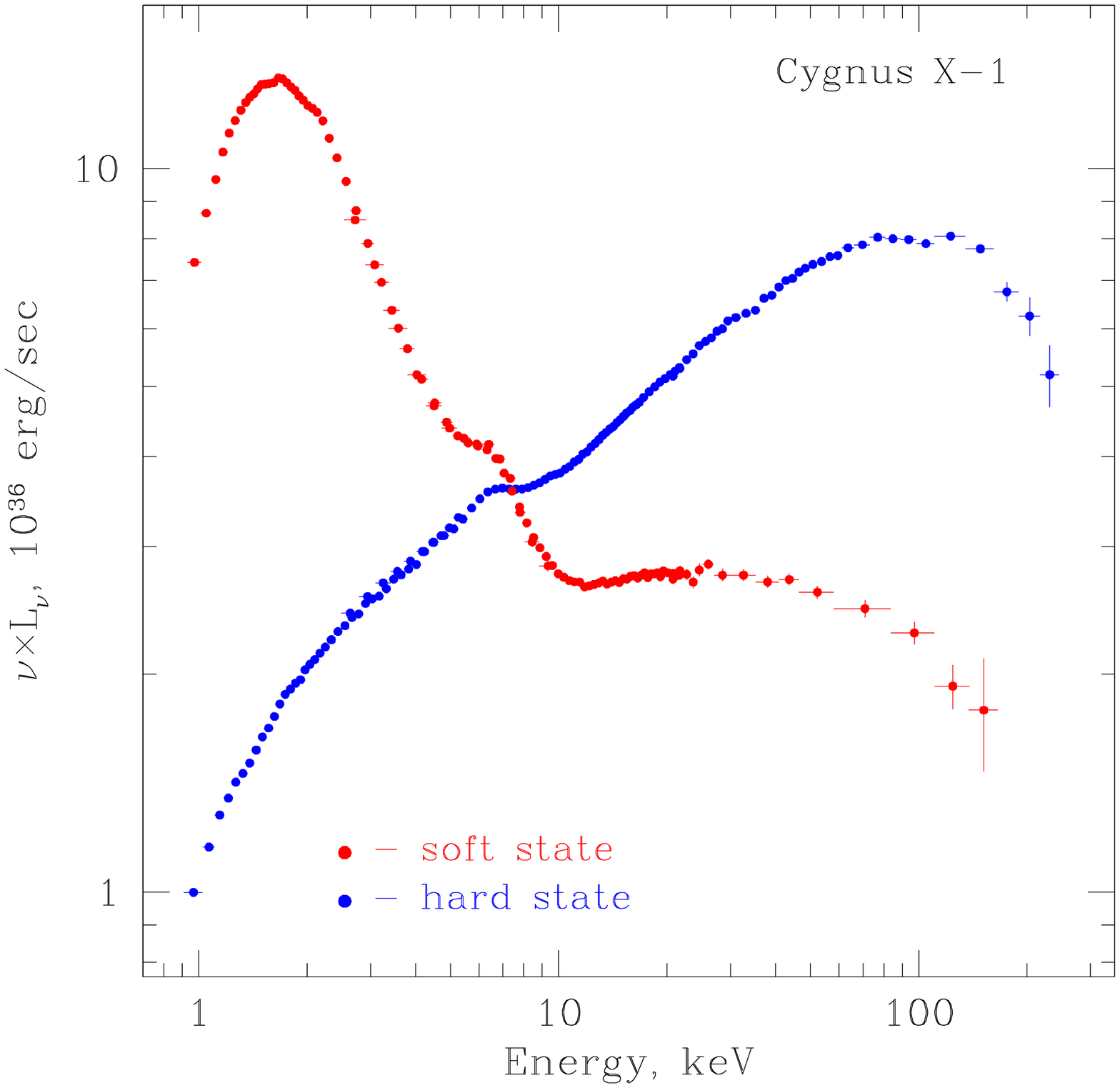}
\hspace{0.2cm}
\includegraphics[width=0.48\textwidth]{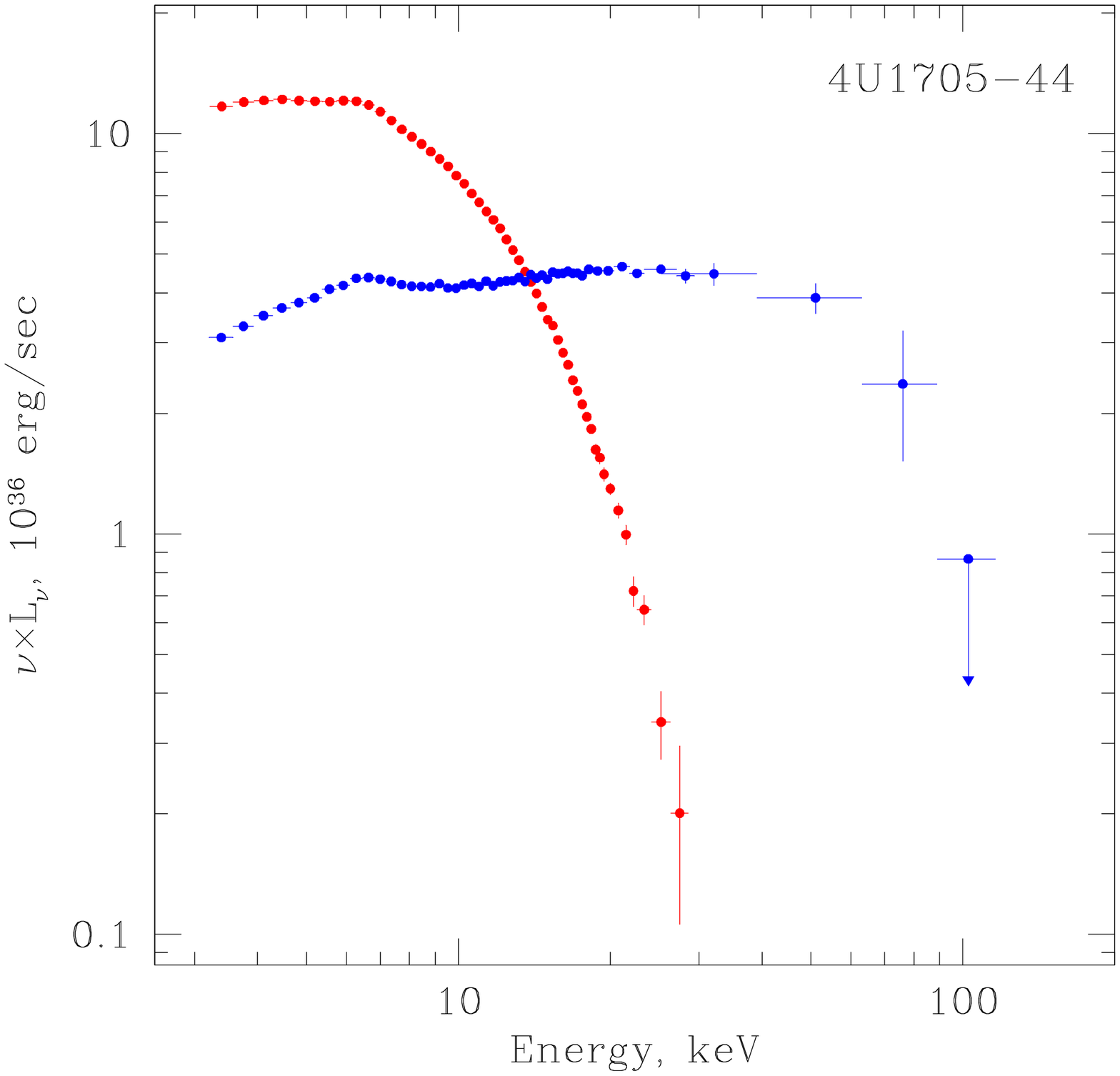}}
\caption{Energy spectra of a black hole Cyg~X-1\index{Cyg~X-1} (left) and a neutron star 4U~1705-44\index{4U~1705-44} (right)  in
the soft and hard spectral states (left panel adopted from \cite{freqres_cygx1}).}
\label{fig:states_spe}
\end{figure}

\section{Geometry and spectral components}
\label{sec:geometry}

The contributions of optically thick and optically thin emission mechanisms can be easily identified in the observed spectra of accreting black holes as soft and hard spectral components (Fig. \ref{fig:states_spe}).  Depending on the spectral state of the source one of these components may dominate the spectrum or they can coexist giving comparable contribution to the total emission.
The soft component is  believed to originate in the geometrically thin and optically thick accretion disk of the Shakura-Sunyaev\index{Shakura-Sunyaev disk} type \cite{ss73}. To zero-th order approximation,
its spectrum can be regarded as a superposition of black body
spectra with the temperatures and emitting areas distributed
according to the energy release and the temperature profile of the
accretion disk. The simplest example of this is the so-called
``multicolor disk blackbody  model'' introduced by \cite{diskbb}. 
Although valid for one particular inner torque boundary condition allowing easy integration of the total flux, this model has been widely and efficiently used in the era of more limited computer resources due to its simplicity, speed and early integration in the XSPEC spectral fitting package.
To achieve a higher degree of accuracy one would need to consider more realistic inner boundary conditions, deviations of the gravitational potential from the Newtonian and to account for such effects as distortion of the black body spectrum due to Thomson scatterings in the upper layers and in the atmosphere of the disk,  Doppler  effect due to rotation  of the matter in the disk etc. A number of models has been proposed to include these effects \cite{grad,shimura_takahara95,ross96,diskns}, many of them currently
implemented in the XSPEC package.

The inadequacy of the optically thick emission mechanism in describing the hard spectral  component present in both spectra in Fig. \ref{fig:states_spe}  can be easily demonstrated.
Indeed, the size of the  emitting region required to achieve a luminosity of $\log(L_X)\sim
37-38$ with black body temperature of $\sim$ several tens of keV is
$R_{em}\sim\left(L_X/\pi\sigma_{SB}T^4\right)^{1/2}
\sim 20\,L_{37}^{1/2}/T_{30}^2$ meters ($T_{30}=T_{bb}/30$ keV). 
Obviously this is much smaller than the size of the region of main energy release near an
accreting black hole, $\sim 50 r_g\sim 500-1500$ km.
The bremsstrahlung
emission from a $\sim$ uniform cloud of hot plasma  
with a filling factor close to unity can not deliver the required
luminosity either, because the $\propto N_e^2$ density dependence of its emissivity makes it very inefficient in the low density regime characteristic of the optically thin accretion flow. Indeed, the optical depth of such a
plasma cloud of size $\sim10-100r_g$ and emission measure
of $N_e^2V\sim 10^{59}$ cm$^{-3}$ required to explain observed hard
X-ray luminosity of $\sim10^{37}$ erg/s would greatly exceed unity. 
However, bremsstrahlung emission may play a role in the advection dominated accretion flow \cite{adaf} in the low $\dot{M}$ regime of quiescent state of accreting black holes ($\log(L_X)\le 32-33$). 

It has been understood early enough that Comptonization is the most plausible process of formation
of the hard spectral component \cite{str79,st80}. 
Thanks to its linear dependence on the gas density,  Comptonization of soft photons on hot electrons in the vicinity of the compact object can efficiently radiate away the energy dissipated in the optically thin accretion flow and  successfully explain the luminosity and overall spectral energy distribution observed in the hard X-ray band. 
Moreover, Comptonization models of
varying degree of complexity do satisfactorily describe the observed
broad band energy spectra of black holes to the finest detail  (e.g. \cite{gx339_osse}).  It is especially remarkable
given the high quality of the X-ray data which became available in the
last decade from observations of recent and current X-ray
observatories, such as Compton GRO, RossiXTE and Chandra, XMM and
Swift. 

\begin{figure}
\centering
\vspace{0.6cm}\includegraphics[width=0.6\hsize]{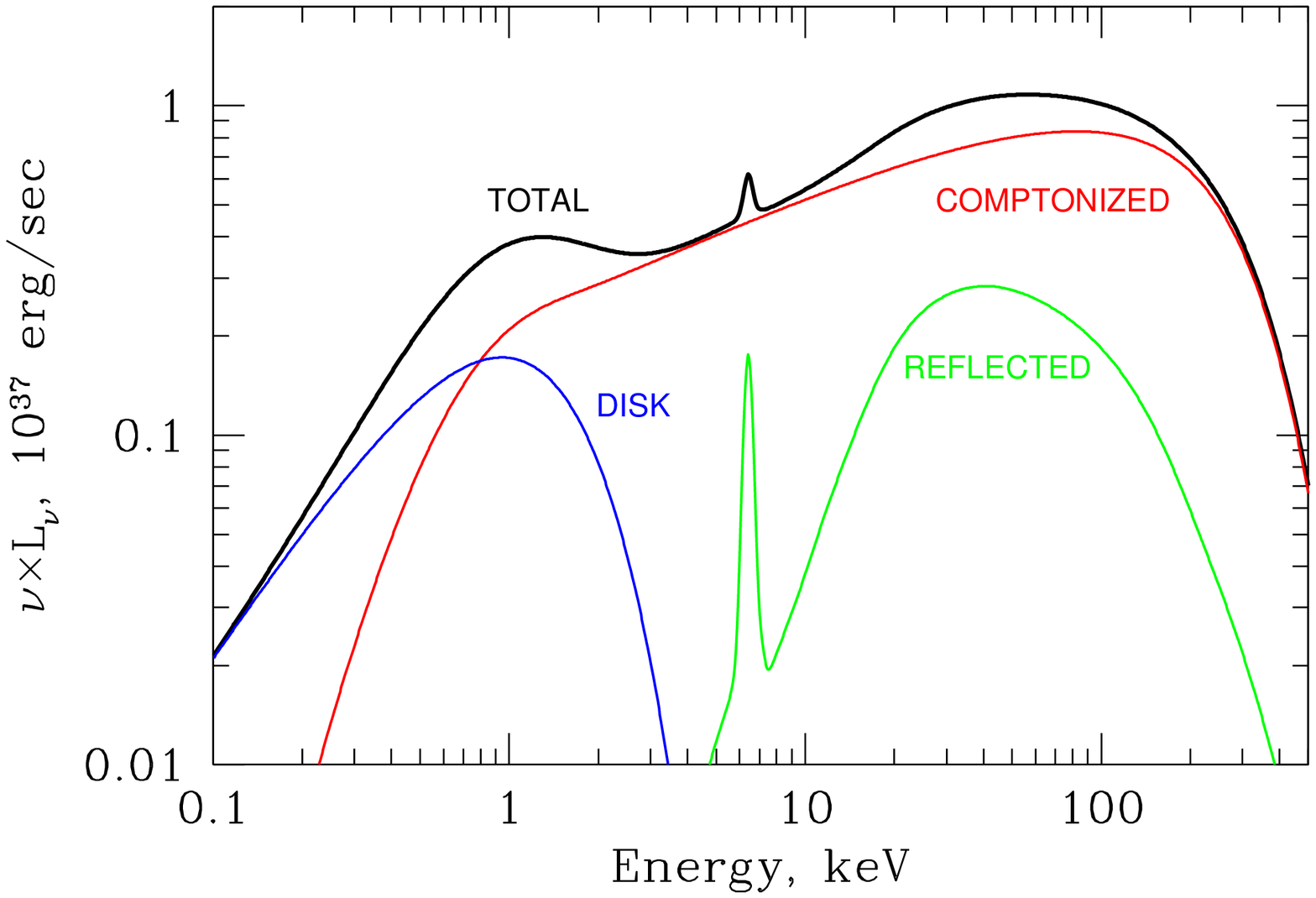}\vspace{0.4cm}
\hbox{\hspace{0.9cm}
\includegraphics[width=0.68\hsize]{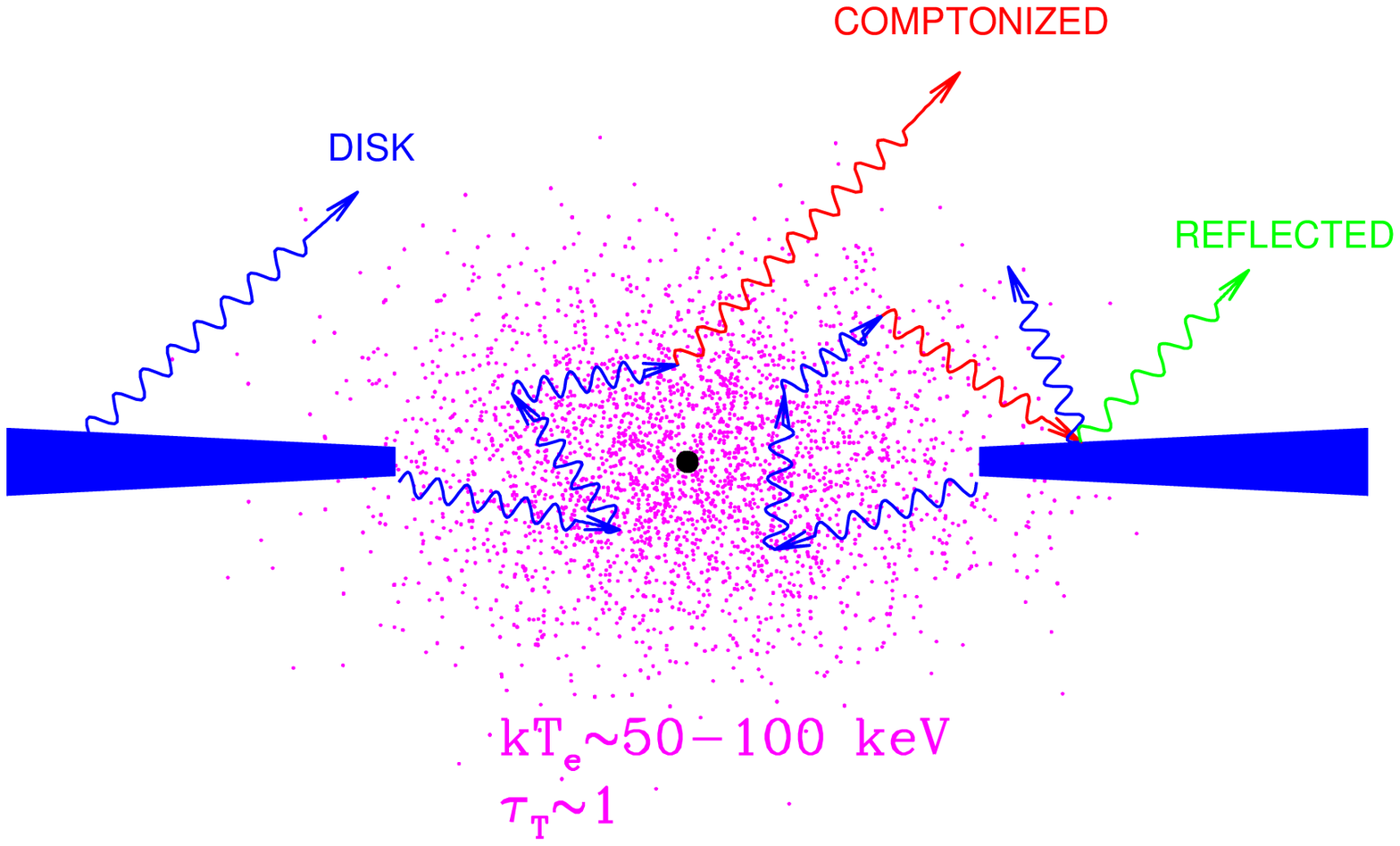}}
\vspace{0.2cm}
\caption{The three main components of the X-ray emission from an accreting black hole  (top) and  a plausible geometry of the accretion flow in the hard spectral state (bottom).
}
\label{fig:geometry}
\end{figure}

The Comptonization site -- cloud of hot (thermal or non-thermal)
electrons is often referred to as a ``corona''.
Although it is generally accepted that the Comptonizing corona has to 
be located in the close vicinity of the compact object, there is
currently no broad consensus on the detailed geometry of the
region. The numerical simulations have not reached the degree of
sophistication required to perform full self-consistent global simulations of
the accretion flow.  Two physically motivated geometries are commonly
considered: (i) the ``sombrero'' configuration (to my knowledge first
introduced in \cite{sombrero}) and (ii) the patchy, flaring corona
above the accretion disk. 

One of the variants of the ``sombrero''
configuration is depicted in Fig.\ref{fig:geometry}. It is assumed that
outside some truncation radius the accretion takes place predominantly via
the optically thick and geometrically thin accretion disk, whereas
closer to the compact object the accretion disk is transformed into a hot
optically thin and geometrically thick flow with the aspect ratio of
$H/R\sim 0.5-1$. The soft (optically thick) and hard (optically thin) spectral components are formed in
the accretion disk and the hot inner flow correspondingly. The
value of the truncation radius can be inferred from the
observations. Although their interpretation is not unique and
unambiguous,  the plausible range 
of values is between $\sim$ 3 and a few hundred gravitational radii. 
There is no  commonly accepted mechanism of truncation of the disk and formation of the 
corona, with a number of plausible scenarios having been investigated
recently. Among the more promising ones is the evaporation
of the accretion disk under the effect of the heat conduction. It was initially suggested to explain quiescent X-ray emission from cataclysmic variables \cite{evap} and was later applied to the case of  accretion onto black holes and neutron stars \cite{evap2}.  It not only provides a physically motivated picture  describing  formation of the corona and destruction of the optically thick disk but also correctly predicts the ordering of spectral states\index{states} vs. the mass accretion rate. Namely, it explains the fact that hard spectra indicating prevalence of the hot optically thin flow are associated with lower $\dot{M}$ values, whereas  the optically thick disk appears to dominate the photon production in the accretion flow at higher  $\dot{M}$. As a historical side note I mention that in the earlier years of X-ray astronomy the presence of the hot optically thin plasma in the vicinity of the compact object  was often associated with disk instabilities, therefore such a behavior appeared puzzling to many astrophysicists in the view of the  $\dot{M}$ dependence of these instabilities.

Another geometrical configuration considered in the context of hard X-ray
emission from black holes is a non-stationary and non-uniform
(patchy) corona above the optically thick accretion disk. 
This scenario has been largely  inspired by the suggestion by Galeev, Rosner and Vaiana \cite{galeev},
that a magnetic field amplified in the hot inner disk by turbulence and differential rotation  may reach  the equipartition value and emerge from the disk in the  form of buoyant loop-like structures of solar type above its surface. These structures may lead to the formation of a hot magnetically confined structured corona similar to the solar corona, which may produce hard emission via inverse Compton and bremsstrahlung mechanisms. The model could also explain the faintness of the hard emission in the soft state as a result of efficient cooling  of plasma in the magnetic loops via inverse Compton effect due to increased flux of soft photons at higher $\dot{M}$.

\begin{figure}
\vspace{1cm}
\centering
\includegraphics[width=0.6\hsize]{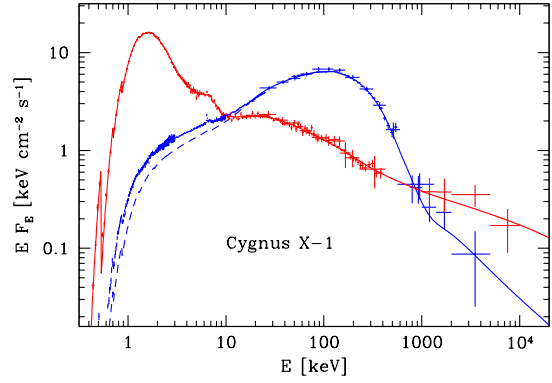}
\caption{Broad band spectra of Cyg~X-1\index{Cyg~X-1} in the soft and hard spectral state based on the data of BeppoSAX and Compton GRO missions. Adopted from \cite{mcconnell}.}
\label{fig:cygx1_bband}
\end{figure}

\begin{figure}
\vspace{1cm}
\centering
\hbox{
\includegraphics[width=0.5\hsize]{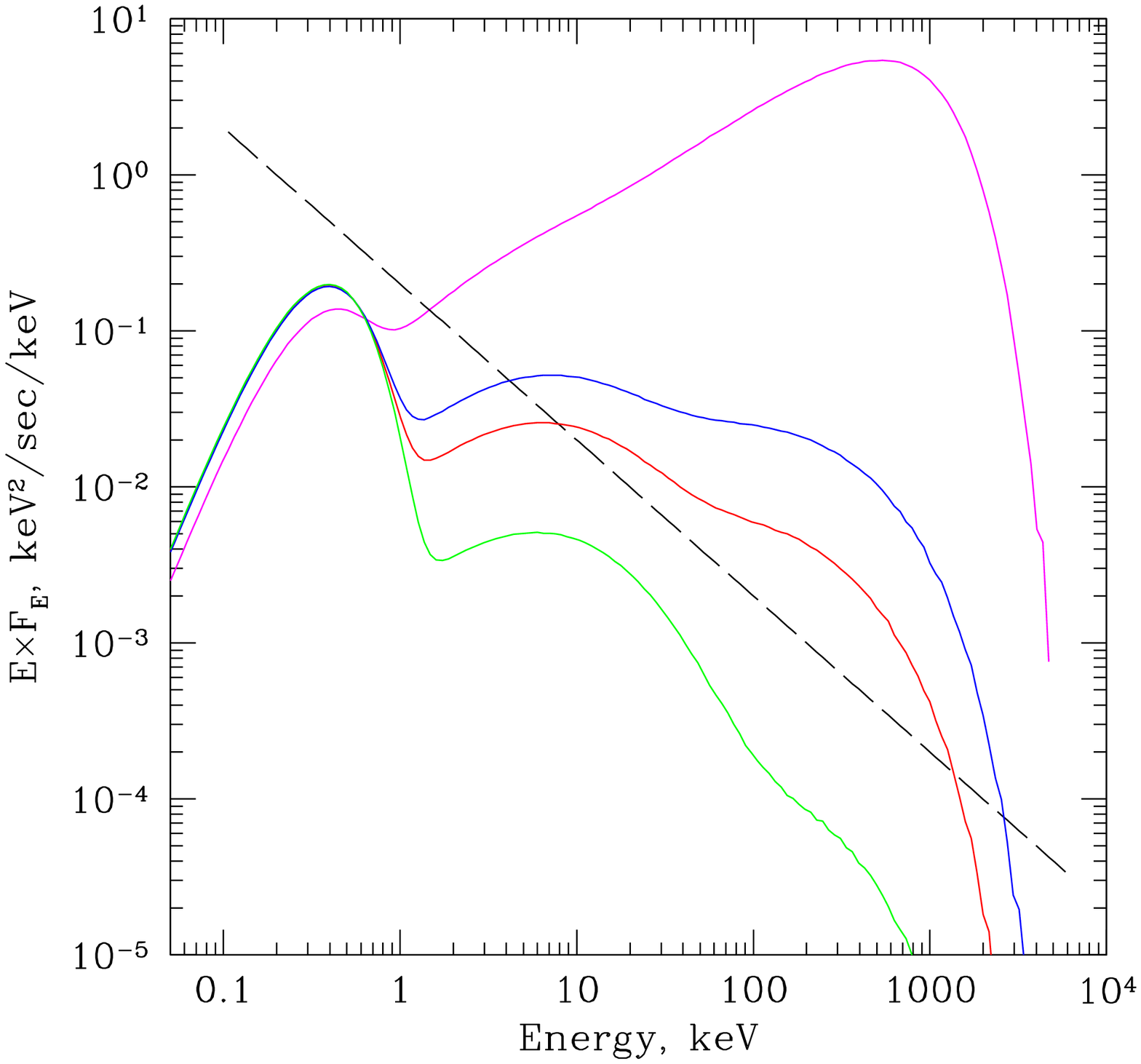}
\includegraphics[width=0.5\hsize]{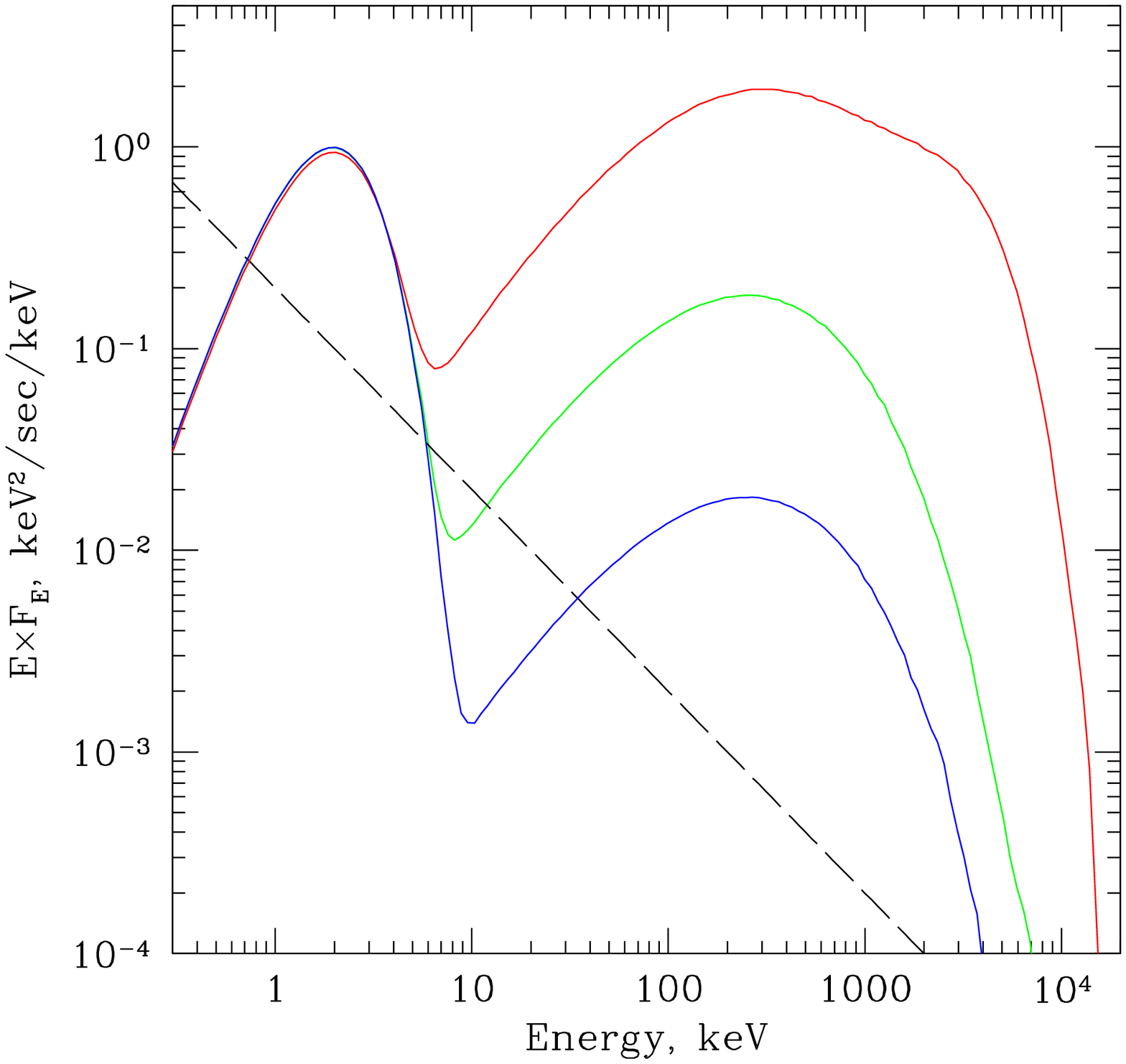}
}
\caption{Comptonization spectrum in the case of large temperature and small optical depth. Left panel:  
$kT_{bb}=100$ eV, $kT_e=300$ keV, $\tau=0.01, 0.05,0.1$ from the bottom to the top; right panel: 
$kT_{bb}=500$ eV, $kT_e=1$ MeV, $\tau=0.001, 0.01, 0.1$. For comparison, the top most spectrum on the left panel shows case of $\tau=1$. The left-most peak in all spectra is made of seed photons  which left the Comptonization region without scatterings. The dashed line is a power law with a photon index of $\Gamma=3$. These spectra are the results of  Monte-Carlo simulations assuming spherical geometry.}
\label{fig:low_tau}
\end{figure}

Although the original Galeev, Rosner and Vaiana paper   was focused on the  thermal emission from hot plasma confined in buoyant  magnetic loops, the latter are a plausible site of particle acceleration responsible for the non-thermal component in the electron distribution. This is expected on theoretical grounds and is illustrated very well observationally by the presence of the non-thermal emission component in the spectra of solar flares (see e.g. \cite{eqpair} and references therein). On the other hand, a power law -shaped  hard X-ray component is commonly observed in the spectra of black hole candidates (Fig.\ref{fig:cygx1_bband}). It reveals itself most graphically in the soft spectral state but may be as well present in the hard state, along with the thermal Comptonized spectrum (e.g.\cite{mcconnell}).  This power law component has a photon index of $\Gamma\sim 2-3$ and, although it may extend into the several hundred keV -- MeV range (e.g. \cite{musca}), it  is relatively unimportant energetically, contributing a small fraction to the total radiation output of the black hole.  This is in contrast with the hard spectral component due to thermal Comptonization, observed in the hard spectral state which accounts for the most of the source luminosity.

There have been attempts to explain the power law component as a result of Comptonization in a media with Maxwellian distribution of electrons of large temperature and small optical depth. This seems to be unlikely as in this case humps due to individual scattering will be seen in the output spectrum, in contrast with the smooth power law spectra observed. This qualitative consideration is illustrated by results of Monte-Carlo simulations shown in  Fig.\ref{fig:low_tau}.

The sombrero configuration is often  associated with a predominantly
thermal distribution of electrons, whereas the solar-like flares above the accretion disk may be the site of electron acceleration   producing non-thermal electron distributions and
power law-like Comptonization spectra. Thermal Comptonization is
believed to be the main mechanism in the hard spectral state, whereas non-thermal
Comptonization is probaly relevant in the soft  
state (Fig.\ref{fig:cygx1_bband}). On the other hand, observations often indicate presence of both
thermal and non-thermal Comptonization components suggesting that the
two types of corona may co-exist (e.g. \cite{askar}).

Due to heating of the disk by  Comptonized radiation from the corona and soft photon feedback,  the uniform thermal corona above the optically thick accretion disk can not  explain observed hard spectra  with photon index of $\Gamma\sim 1.5-2.0$. 
It has been first proposed by Sunyaev and Titarchuk \cite{st89} that presence of cool optically thick media -- e.g. accretion disk or surface of the neutron star, in the vicinity of the Comptonization region will affect the parameters of the latter and, consequently, the shape of the outgoing Comptonized radiation.  Indeed,  some fraction of the Comptonized radiation will be returned to the accretion disk increasing its temperature and, consequently, the soft photon flux to the Comptonization region. This in turn will increase the cooling rate  and will decrease the electron temperature in the Comptonization region leading to  softer and steeper spectra. I will  further illustrate it by the following simple quantitative consideration.
In the sandwich-like geometry, assuming moderate Thompson optical depth of the corona, $\tau_T\sim 1$, a fraction $f\sim 1/2$ of the Comptonized emission will be returned to the accretion disk. Of this, a fraction of $1-\alpha$ ($\alpha\sim 0.2$ -- albedo) will be absorbed and will contribute to the heating of the accretion disk, adding to its heating due to the gravitational energy release. Ignoring the latter, the luminosity enhancement factor in the Comptonization  region, defined as a ratio of its total luminosity to the luminosity of the seed photons (see more detailed discussion in section \ref{sec:toy_models}) will be 
$A\approx (1-\alpha)^{-1}f^{-1}< 2.5$. As well known \cite{enh}, the luminosity enhancement factor is intimately related to the Comptonization parameter $y$ and the photon index $\Gamma$ of the Comptonized radiation. The above constrain on $A$ implies $\Gamma>2.3$ (see Fig.\ref{fig:g-a}), which is  steeper than the hard state spectra typically observed in black holes. This  conclusion is confirmed by the full treatment of the Comptonization problem in ``sandwich'' geometry \cite{haardt}. In order to produce a harder Comptonized spectrum, the value of the feedback coefficient $f$ needs to be reduced. This is achieved, for example, in a non-uniform, patchy and/or non-stationary corona. Another example suggested by \cite{belob} involves bulk motion of the corona with mildly relativistic velocity away from the disk reducing the feedback coefficient  due to the aberration effect (section \ref{sec:toy_models}). A uniform stationary corona above the accretion disk can, in principle, be responsible for  the steep power law component often detected in the soft state, although a non-thermal electron distribution may be a more plausible explanation, as discussed above.
To conclude, I also note that considerations of a similar kind involving the neutron star surface can explain the fact that the neutron star spectra are typically softer than those of black holes (e.g.\cite{st89}, section \ref{sec:ns}).

\section{Spectral states\index{states} and geometry}
\label{sec:states}

The existence of different spectral states is a distinct feature of
accreting X-ray sources, independently of the nature of the compact object (Fig.\ref{fig:states_spe},\ref{fig:cygx1_bband}).  
Although their phenomenology  is far
richer, for the purpose of this chapter I will restrict myself to
the simple dichotomy between soft (high) and hard (low) spectral
states and refer to the next chapter of this book for a more detailed
discussion.
As no global self-consistent
theory/model of accretion exists, all theories explaining 
spectral states have to retreat to qualitative considerations. These
considerations, although phenomenological in nature, usually are based
on numerous observations of black hole systems, simple theoretical
arguments and some simplified solutions and simulations of the
accretion problem. Described below is a plausible, although neither unique nor unanimously accepted, scenario of this kind based on the ``sombrero" geometry of the accretion flow.
There is a number of cartoons and geometry sketches, illustrating this and other scenarios which I will not repeat here and will refer the interested reader to original work, e.g.\cite{aaz_sketch}

As obvious from Fig.\ref{fig:states_spe}, the spectral states phenomenon is
related to the redistribution of the energy released in the optically
thick and optically thin components of the accretion flow. 
In the sombrero configuration one may associate spectral state transitions with change of the disk truncation radius -- the boundary between the outer optically thick accretion disk and the inner optically thin hot flow.

In the soft (aka disk-dominated) spectral state,  the optically thick
accretion  disk extends close to the compact object, possibly to the
last marginally stable Keplerian orbit ($r=3r_g$ for a Schwarzschild
black hole), leaving no ``room" for the hot optically thin flow. Therefore the major fraction of the accretion energy is emitted in
the optically thick accretion disk giving rise to a soft spectrum of the multicolor blackbody type. The magnetic activity at the disk surface may (or may not) produce a hard power-law like tail due to
non-thermal Comptonization in the corona. As discussed in the previous section, this power law component  has a steep slope $\Gamma\sim 2-3$ and is relatively insignificant
energetically.

In the hard (aka corona-dominated) spectral state, the accretion disk truncates at a distance
of $\sim 50-100 r_g$ or further from the compact object. The major fraction of the gravitational energy is released  in the
hot inner flow. Comptonization of the soft photons emitted by the
accretion disk on the hot thermal electrons of the inner flow leads to the
formation of the hard spectrum of the shape characteristic for  
unsaturated thermal Comptonization. The typical parameters in the
Comptonization region -- hot inner flow, inferred from observations
are: electron temperature of $T_e\sim 100$ keV and Thompson optical
depth of $\tau_e\sim 1$.   The significance of the soft backbody-like emission from the optically thick disk as well as of the non-thermal emission due to magnetic flares at its surface
varies depending on the disk truncation radius, increasing as the disk moves inwards. There is evidence that both thermal and non-thermal hard components may co-exist in the hard state in the certain range of the disk truncation radii, as suggested in \cite{askar}.

\section{Reflected emission}
\label{sec:refl}

\begin{figure}
\centering
\includegraphics[width=0.7\textwidth]{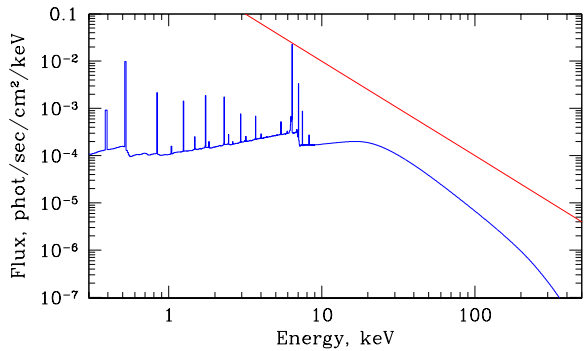}
\caption{Spectrum of emission reflected from an optically thick slab
of neutral matter of solar abundance. The solid line shows the
spectrum of the incident power law emission.}
\label{fig:reflected_mc}
\end{figure}

After escaping the corona, a fraction of the Comptonized photons  may be
intercepted by the optically thick accretion disk. Part  of the
intercepted radiation will be dissipated in the disk, via inverse
Compton-effect and photoabsorption on heavy elements, contributing to
its energy balance, the remaining part will be reflected due to Compton
scatterings \cite{bst}. The disk albedo depends on the photon wavelength 
and on the  chemical composition and ionization state of the disk material. Expressed in terms of energy
flux it is $\sim 0.1-0.2$ in the case of neutral matter of solar
abundance \cite{aaz_albedo}. Some fraction of the photobsorbed emission will be re-emitted
in the fluorescent lines of heavy elements and may also escape the disk. Thus, combined effects of photoabsorption, fluorescence and
Compton scattering form the 
complex spectrum of the reflected emission, consisting of a
number of fluorescent lines and K-edges of cosmically abundant
elements superimposed on the broad Compton reflection hump
\cite{bst,fab} (Fig. \ref{fig:reflected_mc}). 
The peak energy of the latter depends on the shape of
the incident continuum and for a typical spectrum of a black hole in the
hard state is located at $\sim 30$ keV
(Fig. \ref{fig:geometry}).

Due to the dependence of the photoabsorption cross-section, element
abundances and fluorescent yield on the atomic number, the strongest among the spectral
features associated with heavy elements are the K-edge and fluorescent
K-$\alpha$ line of iron. For reflection from the neutral media, the
centroid of the iron line is at 6.4 keV and its equivalent width
computed  with respect to the pure reflected continuum\index{Compton reflection} is $\sim 1$
keV. 
For a distant observer, the reflected emission is diluted with the
primary Comptonized radiation and thermal emission of the accretion disk,
adding complexity to the  observed spectra. 
It leads to the appearance of characteristic reflection features in the
spectra of X-ray  binaries (Fig. \ref{fig:reflected_mc})  -- 
the fluorescent K$_{\alpha}$ line from iron, iron K-edge  and a broad
Compton reflection hump  at higher energies \cite{bst,fab}.  
Their shape is further modified, depending on the ionization state of
the matter in the disk \cite{ionrefl}, by strong gravity effects  
and Doppler and aberration  effects due to the Keplerian rotation of the disk
e.g. \cite{fab1}. The observed line usually has non-zero intrinsic
width and the K-edge is never sharp (smeared edge).
Their amplitude depends on the ionization state and on the solid 
angle of the reflector as seen from the source of the primary
radiation. This can change with the spectral state of the source. The
typical value of the iron line equivalent width is $\sim 50-300$ eV.

The presence of theses features makes the spectra deviate
from a power law shape, expected for  Comptonized radiation with
parameters typical for black holes. As their amplitude depends on the
relative configurations of the corona and the accretion disk and
physical conditions in the disk, they have great diagnostics potential
for studying the accretion geometry in different spectral states. This is further discussed in Sect. \ref{sec:correl}. 

\section{Polarization of X-ray emission}

As the distribution of matter in the accretion flow is not spherically symmetric, one may expect some degree of polarization\index{polarization} of emerging X-ray radiation. In the conventional scenario of formation of X-ray radiation in the vicinity of a compact object the polarization is caused  by Thompson scatterings of photons on free electrons in the accretion disk and hot corona and is predicted to be present at the moderate level of $\sim$ several percent  in all three main components of the X-ray emission: thermal emission of the optically thick disk, Comptonized emission and reflected component. 
A considerably larger degree of polarization may be expected in some alternative scenarios, for example in some versions of the jet scenario, discussed later in this book.

The degree of
polarization of the Comptonized emission depends strongly on the
geometry of the corona, the location of the sources of the soft seed
radiation and the viewing angle \cite{st_polariz, polariz_pulsars}. 
As calculated by Sunyaev \& Titarchuk \cite{st_polariz}, for the case
of Comptonization in the disk it lies in the range between 0 and
$\sim 12\%$, depending on the optical depth of the corona and
viewing angle, i.e. it can slightly exceed the maximum value of 11.7\%
for a pure-scattering semi-infinite  atmosphere. Although the disk
geometry of the corona is unlikely, this result suggests that a moderate
degree of polarization of the Comptonized continuum, of the order of
$\sim$few per cent, may be expected. 

Obviously, polarization\index{polarization} should also be expected for the reflected emission
from the accretion disk. The pure reflected component can be
polarized to 
$\sim 30\%$ \cite{polariz1}. The degree of polarization drops to
$\sim$ few per cent when it is diluted with the Comptonized radiation
and the thermal emission from the disk. The degree of polarization is
a strong function of photon energy -- it is low at low energy
$< 10$ keV and reaches its maximum of $\sim 5\%$ (for a 60 degree
viewing angle) at $\sim 30-50$ keV,
where the contribution of the Compton-reflected continuum\index{Compton reflection} to the
overall spectrum is maximal \cite{polariz1}. 

The thermal emission generated inside the optically thick accretion disk is initially unpolarized but attains polarization as a result of Thompson scatterings in the atmosphere of the accretion disk \cite{connors,lixin}. Calculations \cite{lixin} show that the degree of polarization varies from 0\% (face-on disk) to $\sim 5\%$ (edge-on).
The polarization degree depends on the photon energy, being largest in the Wien part of the thermal spectrum, where scatterings play the most important role.
A remarkable property of polarized radiation  from the vicinity of
a compact object is that due to relativistic effects the polarization
angle also becomes dependent on the photon energy \cite{connors,lixin}. 
This is in contrast with the classical approximation, where
symmetry considerations require that the polarization direction is
coaligned with the minor or major axis of the disk projection on the
plane of the sky. 
The amplitude and shape of these dependences are sensitive to the disk inclination and the spin of the black hole. In combination with spectral information, this can be used to resolve the degeneracy between the black mass, spin and disk inclination \cite{lixin}. 

Polarization measurements are yet an unexplored area of high energy 
astrophysics and have great diagnostics potential. With the new generation of polarimetric detectors,  
X-ray polarimetry will become  a powerful tool to study the geometry of the
accretion flow and the properties of the compact object in accreting systems. 

\section{Variability}
\label{sec:variability}

Variability of X-ray emission  is  a common and well-known property of X-ray binaries. The amplitude of flux variations depends on the time scale and photon energy and can be as large as $\sim 20-30\%$ fractional rms. The dependence of the variability amplitude on  time scale is conventionally characterized by the power density spectrum, which is a square of the Fourier-amplitudes of the time series, renormalized in order to give the answer in desired units, for example (fractional rms)$^2$/Hz.  
Power density spectra of X-ray sources  often reveal a number of  rather narrow features of various widths superimposed on a broad band continuum of aperiodic variations, suggesting that both resonances of various degree of coherence as well as stochastic processes of aperiodic nature contribute to the observed flux variations. Some of the narrow features can be clearly associated  with the spin frequency of the neutron star or orbital or precession frequency of the binary. The nature of others, broadly reffered to as quasi-periodic oscillations (QPO\index{QPO}), is poorly understood.  The models range  from the beat frequency model  proposed by A.Alpar and J.Shaham \cite{alpar85} soon after the discovery of the first QPOs in neutron stars, and the relativistic precession  model \cite{stella} relating QPO frequencies with fundamental  frequencies  of Keplerian orbits in strong gravity to models employing global oscillation modes in the accretion disk \cite{abr_qpo, tit_qpo}. QPOs were first discovered in neutron star systems, but are also commonly observed in accreting black holes. Their rich phenomenology is well documented elsewhere  (e.g.\cite{qporev}) and is not discussed in this chapter, which will focus on the aperiodic variability continuum. 

I will conclude these introductory remarks with the following side note. The interpretation of the energy spectra of celestial X-ray sources has been greatly facilitated by the fact that a number of simple physical concepts could be employed in a straightforward manner, such as black-body emission, Comptonization, photoabsorption and fluorescence etc. 
No equivalent concepts are easily available to help in interpreting the power density spectra. This may explain why our understanding of the power density spectra lags significantly behind our understanding of the energy spectra. Indeed, spectral analysis can rely on a number of physically motivated and elaborate models which successfully describe high quality data from modern observatories.  The interpretation of the power density spectra, on the contrary,   has just started to  advance beyond ``numerology" and simple association of QPO\index{QPO} frequencies with characteristic frequencies\index{characteristic frequencies} of test particles in the (strong) gravitational field. However, it is obvious that timing information (along with the polarization measurements) presents a completely different dimension  which has to be taken into account in validating any model of formation of radiation in the vicinity of a compact object.  
\bigskip

\begin{figure}
\centerline{    
\includegraphics[width=0.5\hsize]{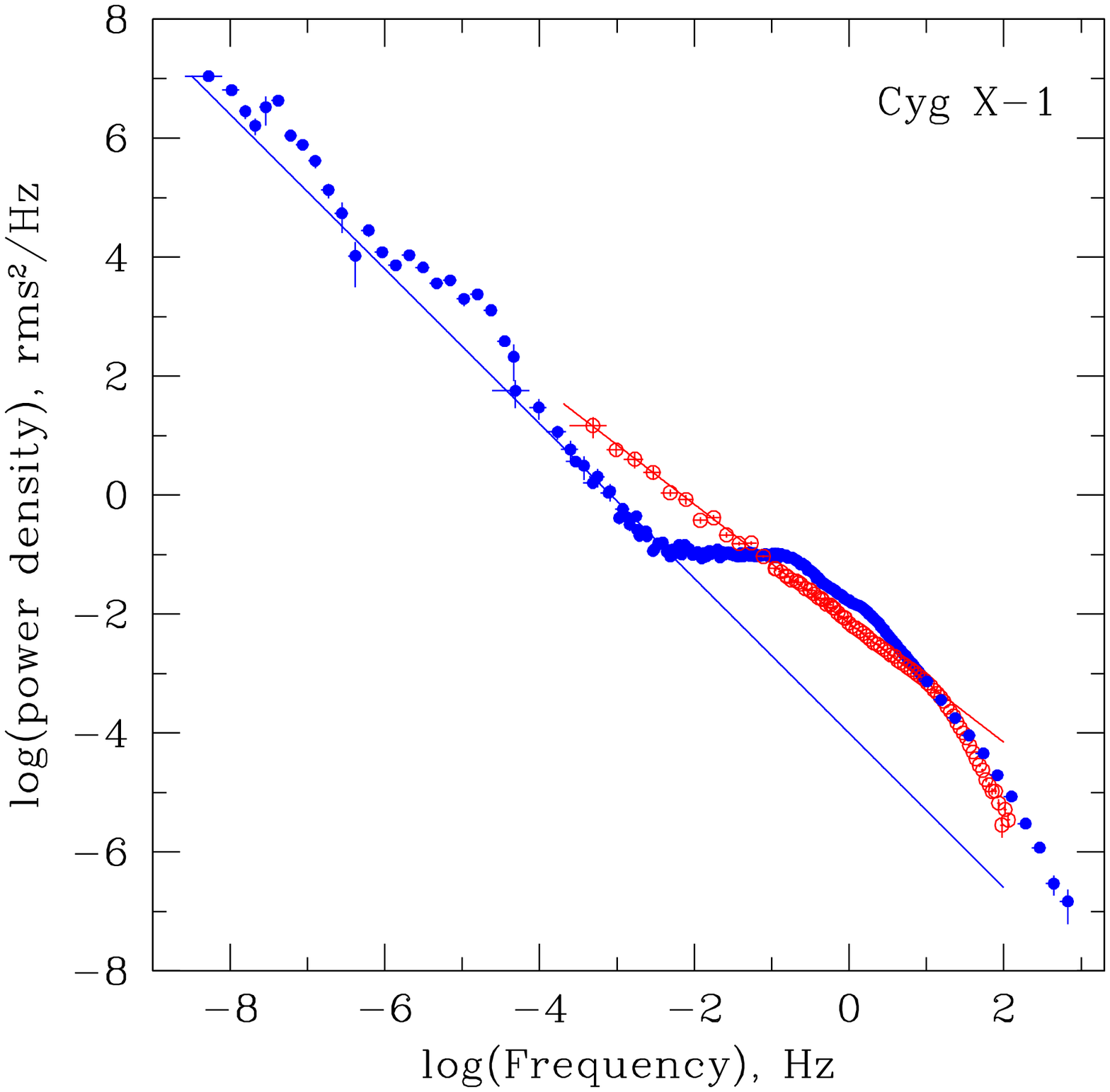}
\includegraphics[width=0.5\hsize]{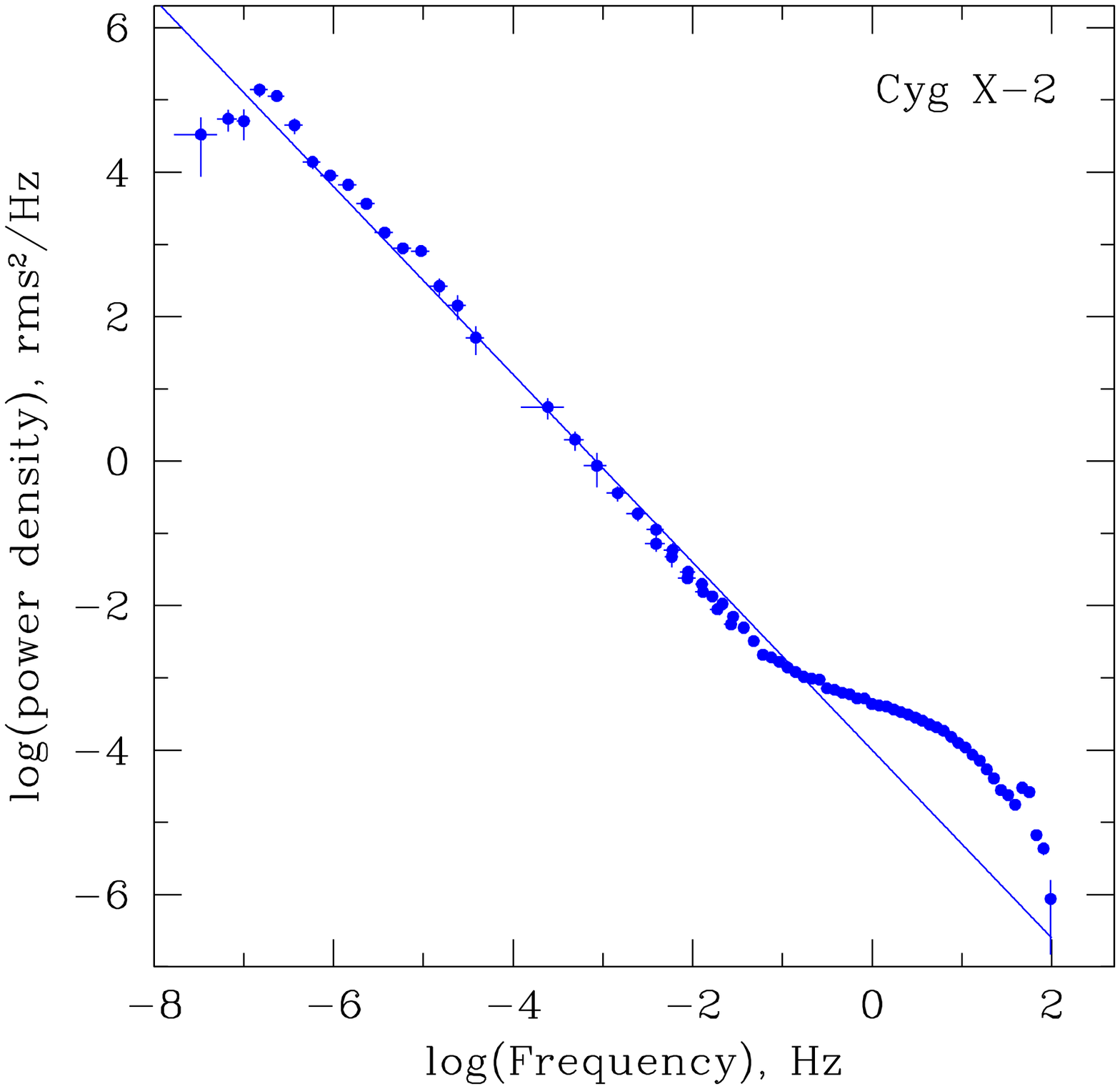}
}
\caption{Broad-band power density spectra of Cyg~X-1 \index{Cyg~X-1} and Cyg~X-2\index{Cyg~X-2}. 
The spectra are obtained from non-simultaneous  RXTE/ASM, EXOSAT/ME and RXTE/PCA data.
In the left panel, the open circles at $\log(f)>-4$ (red in the color version of the plot) show the 1996
high state PDS, the filled circles (blue)  are hard
spectral state PDS. 
The solid lines extending through entire plots in the left and right panels show  a $P_\nu\propto\nu^{-1.3}$ 
power law;  normalization is the same on both panels. The second power law in the left panel giving a better approximation to the soft state data of Cyg~X-1 \index{Cyg~X-1} is $P_\nu\propto\nu^{-1}$ power law.
The very high frequency data
$\log(f)>1$ are from \cite{cygx1_highfreq}. 
} 
\label{fig:pds_broadband}
\end{figure}

\subsection{Propagation of $\dot{M}$ fluctuations in the accretion disk}
\label{sec:fluc}

The remarkable characteristics of aperiodic variability is its
breadth in the frequency domain. As illustrated by the power density
spectra of Cyg~X-1 \index{Cyg~X-1} and Cyg~X-2 \index{Cyg~X-2} shown in Fig. \ref{fig:pds_broadband}, flux variations in
the frequency range from $\log f<-8$ to $\log f\sim 2-3$ are present. 
This suggests that variations of the mass accretion rate $\dot{M}$ on
an extremely broad range of time scales are present in the innermost
region of the accretion flow, $r\le50-100\,r_g$, where X-ray emission is
produced. It is to be compared with characteristic time scales in this region which are limited by two extremes --  the dynamical time scale that is of the order of the Keplerian orbital time 
$t_K\approx 0.3\left(M/10 M_\odot\right)\left(r/50r_g\right)^{3/2}$ sec
and the viscous time scale 
$t_{visc}\sim \alpha^{-1} \left(h/r\right)^{-2} \Omega_K^{-1}= (2\pi\alpha)^{-1}\left(h/r\right)^{-2}t_K$ ($\alpha$ is viscosity parameter, $h$ -- disk thickness). The latter is in the range between 
$t_{visc}\sim 10^4\,t_K\sim 10^3$ sec for a gas pressure dominated Shakura-Sunyaev \index{Shakura-Sunyaev disk}  disk ($h/r\sim 10^{-2}$) and $t_{visc}\sim 10\,t_K\sim 1-10$ sec for a thicker hot flow with the aspect ratio of $h/r\sim 1/3$.

From the point of view of characteristic time scales,  the high-frequency  variations could potentially be
produced  in the vicinity of the compact object. Longer time scales, on the contrary,  exceed by many orders of magnitude  the longest time scales in the region of the main energy release and can not be generated there.  The low frequency $\dot{M}$ variations have to be generated in the outer parts of the accretion flow and be propagated to the region of the main energy release where they are converted into variations of the X-ray flux. The power spectra shown in  Fig. \ref{fig:pds_broadband} maintain the same power law shape over a broad frequency range, suggesting that the same physical mechanism is responsible for flux variations at all frequencies. 
The same slope and normalization of the power law component in the power density spectra of different sources -- black holes and neutron stars -- suggests that this mechanism is a property of the accretion disk and does not  depend on the nature of the compact object. 
A plausible candidate for such mechanism may be viscosity fluctuations caused for example by MHD turbulence in the accretion disk, as suggested in \cite{lyub97}.
Viscosity fluctuations lead to  variations of the  mass accretion rate which after propagating into the innermost region of the accretion flow are transformed into variations of the X-ray flux.

However, because of the diffusive nature of disc accretion \cite{ss73}, fluctuations of mass accretion rate on time scales much shorter than the diffusion time scale will be not propagated inwards, but will  be damped close to the radius at which they originated. As demonstrated in \cite{churazov01}, the amplitude of the fluctuations on the time scale $\tau$ will be significantly suppressed at the characteristic length scale $\Delta r/r\sim \sqrt{\tau/t_{visc}}$. As the viscous time scale depends quadratically on the disk thickness, $t_{visc}\sim \left(h/r\right)^{-2}\Omega_K^{-1}$,  fluctuations on the dynamical $\tau\sim t_d\sim \Omega_K^{-1}$ or thermal $\tau\sim t_{th}\sim \left(\alpha\Omega_K\right)^{-1}$ time scales will be damped  in the thin disk  with $h/r\ll1$ after traveling a small distance in the radial direction, $\Delta r/r\sim h/r$, and will never reach the region of the main energy release. 
The inner region itself can not generate $\dot{M}$ fluctuations on the dynamical or thermal time scales either. Indeed,  the coherence length for fluctuations on a time scale $\tau\sim \Omega_K^{-1}$ is $\Delta r/r\sim h/r\sim 10^{-2}$ for the gas pressure supported Shakura-Sunyaev\index{Shakura-Sunyaev disk} disk. Therefore  $N\sim r/\Delta r\sim 100$ independent annuli regions will contribute to the observed flux, their variations being uncorrelated. The contribution of each region to the total flux is small, $\sim 10^{-2}$, and alone can not cause significant variability of the total flux. Furthermore, in their combined emission uncorrelated fluctuations will be suppressed by a factor of  $\sim 1/\sqrt{N}\sim 10^{-1}$ due to the averaging effect. Thus, for the geometrically thin disc, fluctuations of viscosity (or mass accretion rate) on the dynamical or thermal time-scales will not contribute significantly  to observed  variability of the X-ray flux.
In order to lead to  significant modulation of  the X-ray flux, viscosity or accretion rate fluctuations  have to be propagated inwards significant radial distance and cause fluctuations of mass accretion rate in a significant range of smaller radii, including the region of the main energy release. This is only possible for fluctuations on   time scales equal or longer than the viscous time of the disk at the radius, where fluctuations are ``inserted" into the accretion flow \cite{lyub97,churazov01}.

Thus, the standard Shakura-Sunyaev\index{Shakura-Sunyaev disk} disk plays the role of a low-pass filter, at any given
radius $r$ suppressing $\dot{M}$ variations on time scales shorter than the local viscous  time $t_{\rm visc}(r)$. The viscous time is,  on the other hand, the longest  time scale of the accretion flow, therefore no significant fluctuations at longer time scales can be produced at any given radius. Hence,  a radius $r$  contributes to  $\dot{M}$ and X-ray flux variations predominantly at a frequency $f\sim t_{\rm visc}(r)^{-1}$ \cite{lyub97,churazov01}. The broad range of variability time scales observed in the X-ray power density spectra (Fig.\ref{fig:pds_broadband}) is explained by the broad range of radii at which viscosity fluctuations are generated.  If viscosity fluctuations at all radii have the same relative amplitude, a power law spectrum $P_\nu\propto \nu^{-1}$ will naturally appear \cite{lyub97}, in qualitative  agreement with observations (Fig.\ref{fig:pds_broadband}). The picture of inward-propagating fluctuations outlined above  also successfully explains  the observed linear relation between rms amplitude of aperiodic variability and total X-ray flux in black holes \cite{uttley} and the nearly logarithmic dependence of the time lag between time series in different energy bands on the photon energy \cite{kotov}.

\subsection{Very low frequency break and accretion disk corona}
\label{sec:vlf_break}

\begin{figure}[htbp]
   \centering
   \includegraphics[width=0.7\hsize]{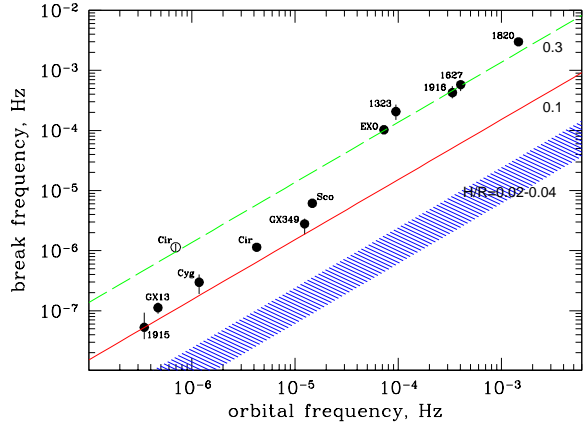} 
   \caption{Relation between the  break frequency ($\sim t_{visc}^{-1}$  at the outer edge of the accretion disk) and the orbital frequency of the binary system. The shaded area shows the  dependence $f_{\rm visc}$ vs. $f_{\rm orb}$ expected for the standard gas pressure supported Shakura-Sunyaev disk ($h/r\sim {\rm few}\times10^{-2}$). Straight solid and dashed lines are predictions for larger values of the disk thickness $h/r$, as indicated by the numbers on the plot. The two points for Cir X-1 correspond to the original data (open circle) and after correction for the eccentricity of the binary orbit  (filled circle). Adopted from \cite{lmxb_tvisc}.}
   \label{fig:fbreak}
\end{figure}

Owing to the finite size of the accretion disk, the longest  time scale in the disk is restricted by the viscous time on its outer boundary $t_{visc}(R_d)$. Below this frequency,  X-ray flux variations are uncorrelated, 
therefore the power density spectrum should become flat at $f\le t_{visc}(R_d )^{-1}$. This explains the low frequency break clearly seen in the power spectrum of Cyg~X-2\index{Cyg~X-2} (Fig. \ref{fig:pds_broadband}). If there are several components in the accretion flow, for example, a geometrically thin disk and a diffuse  corona above it, several breaks can appear in the power spectrum at frequencies corresponding to the inverse viscous time scale of each component. It can be easily demonstrated  \cite{lmxb_tvisc} that the break frequency is related to the orbital frequency of the binary via:
$$
f_{break}=3\pi\alpha (1+q)^{-1/2} \left(h/r\right)^2 \left(R_d/a\right)^{-3/2} f_{orb}
$$
For low-mass X-ray binaries (Roche-lobe filling systems)  this becomes
$f_{break}\propto \left(h/r\right)^2 f_{orb}$.
Such  very low frequency breaks indeed are observed in a statistically representative sample of low mass X-ray binaries, and the break frequency is inversely proportional to the orbital period of the binary (Fig. \ref{fig:fbreak}, \cite{lmxb_tvisc}).

However, measured values of the break frequency imply that the viscous time of the accretion flow is  a factor of $\ge 10$ shorter than predicted by the standard theory of accretion disks (cf. shaded area in Fig. \ref{fig:fbreak}).  This suggests that  significant fraction of the accretion $\dot{M}$ occurs through the geometrically thicker coronal flow above the standard thin  disk. Note that  the existence of the Shakura-Sunyaev \index{Shakura-Sunyaev disk} disk underneath the coronal flow is required by optical and UV observations of LMXBs indicating the presence of the optically thick media of extent  comparable with the Roche-lobe size \cite{lmxb_tvisc}.  The aspect ratio of the coronal flow implied by the $f_{br}/f_{orb}$ measurements, $h/r\sim 0.1$, corresponds to a gas temperature of $T\sim 0.01 \, T_{\rm vir}$.  The corona has moderate optical depth in the radial direction, $\tau_T\sim 1$, and contains $\le 10\%$ of the total mass of the accreting matter (but the fraction of $\dot{M}$ is much larger, probably $\sim 0.5$). These estimates of temperature and density of the corona are in quantitative agreement with the parameters inferred by the X-ray spectroscopic observations by Chandra and XMM-Newton of complex absorption/emission features in LMXBs with large inclination angle.  

In this picture, the red noise\index{red noise} (power law) component of the observed variability of the X-ray flux is defined by the viscosity and/or $\dot{M}$  fluctuations generated  in the coronal flow rather than in the geometrically thin disk. Fluctuations produced in the standard Shakura-Sunyaev\index{Shakura-Sunyaev disk} disk have much smaller amplitude due to a $\sim 10$ times longer viscous time scale (see discussion in Sect. \ref{sec:fluc}) and do not contribute significantly  to the observed variability of X-ray flux. This is further supported by the lack of variability in the soft black body component in Cyg~X-1\index{Cyg~X-1}, as discussed below.

\subsection{High and very high frequencies}
\label{sec:high_freq}

In addition to the power law component of red noise dominating in the larger part of the low frequency range, power density spectra usually have significant excess of power in the high frequency end, $f\ge 10^{-2}-10^{-1}$ Hz (Fig. \ref{fig:pds_broadband}). This excess power is often referred to as a ``band-limited noise\index{band-limited noise}", due its limited scope in frequency.  It is this part of the power spectrum that is the subject of investigation in the majority of ``standard" timing analysis projects, for example in the ones based on a typical  RXTE/PCA observation.

\begin{figure}[htbp]
   \centering
   \includegraphics[width=0.47\hsize]{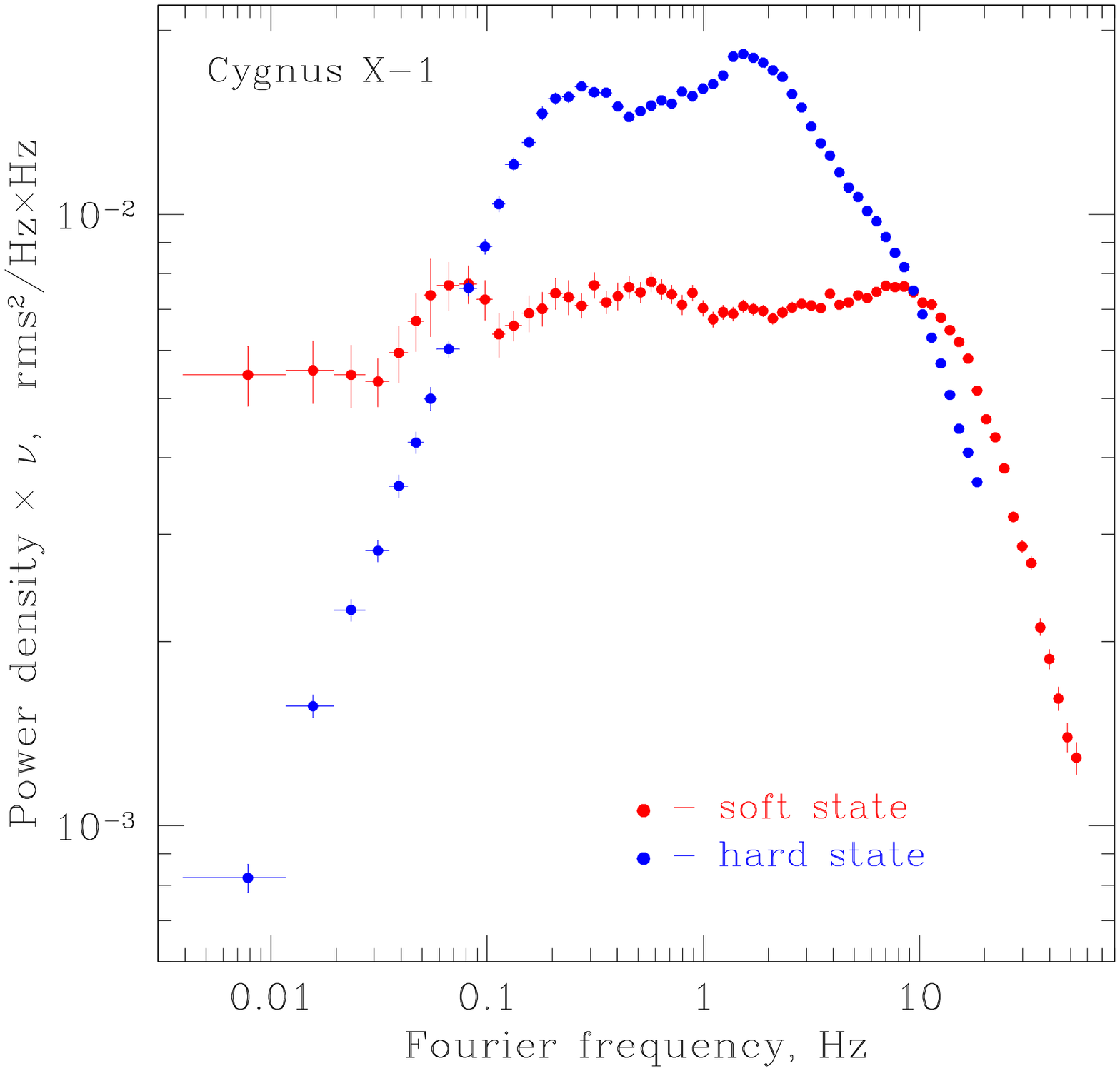} 
    \includegraphics[width=0.50\hsize]{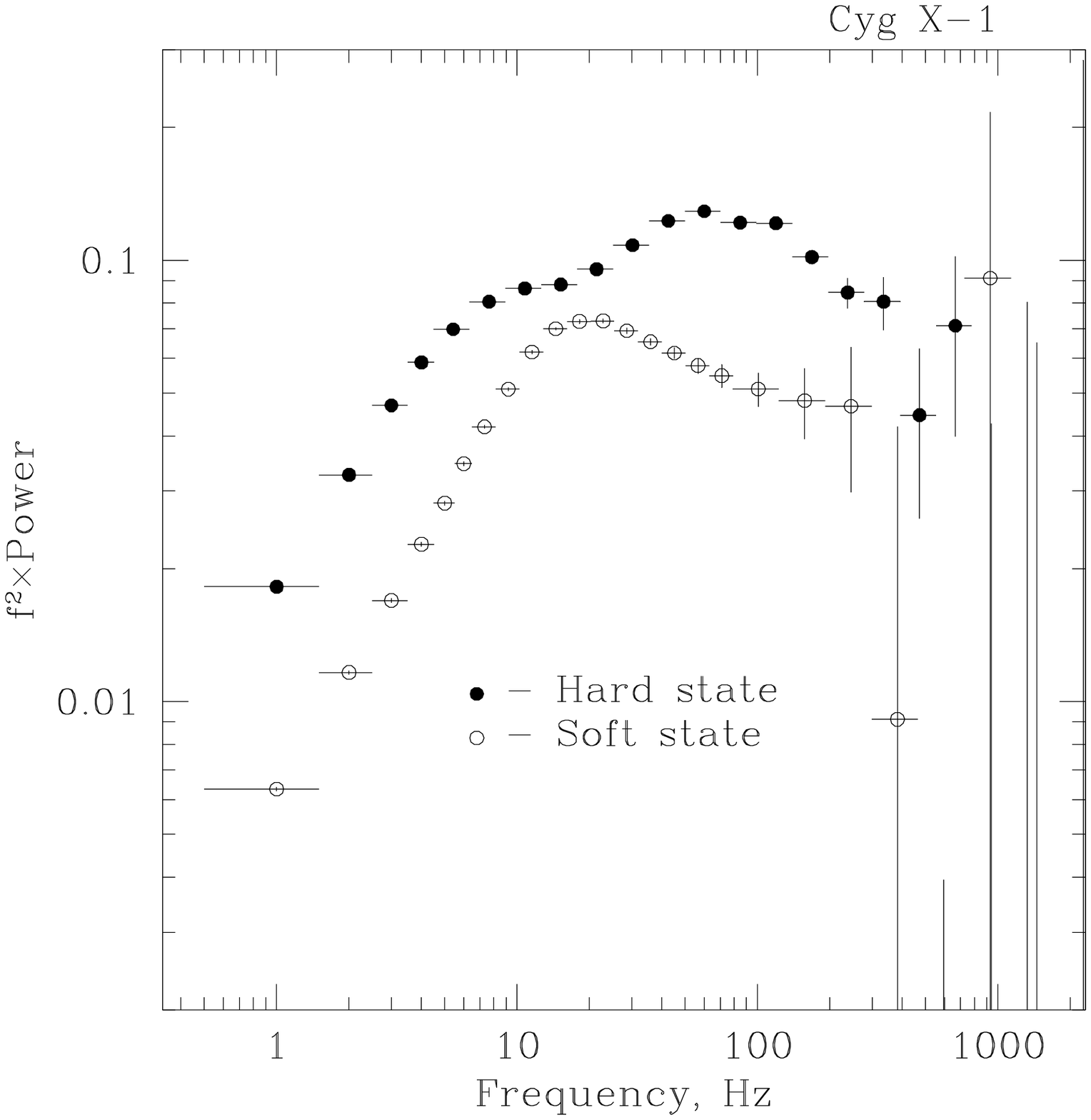} 
   \caption{Power density spectra of Cyg~X-1\index{Cyg~X-1} in the soft and hard spectral states. The right hand panel shows the very high frequency part of the power density spectrum. The power is shown in units of $\nu P_\nu$ (left) and $\nu^2 P_\nu$ (right). Adopted from  \cite{freqres_cygx1} and \cite{cygx1_highfreq} respectively.}
   \label{fig:pds_states}
\end{figure}

In black holes, this component is most prominent in the hard spectral state and usually disappears in the soft state (Fig. \ref{fig:pds_broadband}, \ref{fig:pds_states}). Plotted in the $\nu P_\nu$ units, it shows several broad humps, separated by a $\sim 0.5-1$ decade in frequency (Fig. \ref{fig:pds_states}).  The centroid frequencies of these bumps vary with time but they usually do correlate with each other \cite{paper1,wvdk99}.  It has been proposed that they may be identified with the (much more narrow) QPO\index{QPO} peaks observed in the power spectra of accreting neutron stars. When the proper identifications are made,  black hole and neutron star systems  appear to follow the same global relation between QPO frequencies \cite{pbk,wvdk99,nowak2000,bpk}. This suggests that the features on the power density spectra and their frequencies  are the property of the accretion disk, their existence being unaffected by the nature of the compact object.

Based on the amplitude and timescale considerations, it is plausible to link the origin of the band-limited noise\index{band-limited noise} component to the inner hot optically thin flow in the region of the main energy release. For a $10M_\odot$ black hole, the characteristic time scales in this region ($r\sim 100 r_g$) are: $t_K\sim 1$ sec and $t_{visc}\sim 10$ sec, the latter was computed assuming $\alpha\sim 0.2$ and $h/r\sim 0.2$. This suggest that the two main peaks observed in the $\nu P_\nu$ spectra may (or may not) be related to the dynamical and viscous time scales at the inner edge of the truncated accretion disk and their origin may be linked to the interaction between the geometrically thin outer accretion disk and optically thin hot inner flow.  Note that what appears to be the lower frequency hump in the $\nu P_\nu$ plot  is in fact a break, below which the power density distribution is flat, down to the frequency where the red noise\index{red noise} component becomes dominant (Fig. \ref{fig:pds_broadband}). As a flat power density distribution means lack of correlation between events on the corresponding time scales, we can draw an analogy with the very low frequency breaks associated with the viscous time scale at the outer edge of the disk (Sect. \ref{sec:vlf_break}). This further supports the association of the low frequency hump in the $\nu P_\nu$ plot,  Fig. \ref{fig:pds_states} (= break in the power density spectrum)  with the viscous time scale at the outer boundary of the hot inner flow.  One can estimate the truncation radius of the geometrically thin disk (= outer radius of the hot inner flow in the sombrero configuration, Sect. \ref{sec:geometry}) equating the viscous time with the inverse break frequency, $f_{br}^{-1}\sim 5$ sec: $r_{tr}\sim 10^2 r_g$ assuming $\alpha=0.2$ and the aspect ratio of the hot inner flow $h/r=1/3$. This number is in a good agreement with the disk truncation radius inferred by other measurements, e.g. derived from the variability of the reflected component (\cite{freqres_cygx1}, Sect. \ref{sec:var_refl}).

The right panel in Fig. \ref{fig:pds_states} also demonstrates that there is considerable power at very high frequencies, at the level of  $\nu P_\nu \sim 1-3\%$  at $f\sim 10^2$ Hz. The millisecond range  includes a number of important characteristic time scales in the vicinity of the compact object.  The Keplerian frequency at the last marginally stable orbit for a non-rotating black hole is $f_K\sim 220 \left( M/10M_\odot\right)^{-1}$ Hz. The relativistic precession frequencies associated with Keplerian motion in the vicinity of the compact object are also in this range \cite{stella}. The typical value of the sound crossing time in the gas pressure supported thin disk corresponds to the frequency of a few hundred Hz. The light crossing time for the region of $\sim 10 r_g$ corresponds to the frequency of  $\sim 1$ kHz. However, the observed power spectra break somewhere between $f\sim 20$ Hz (soft state) and $f\sim 50-100$ Hz (hard state) and do not exhibit any detectable features beyond these frequencies. The upper limit on the fractional rms of aperiodic variability in the 500--1000 Hz range is $\approx 1\%$. Similar upper limit can be placed on the rms amplitude of narrow features anywhere in this range, $\approx 0.9\%$ assuming width $\nu/\Delta\nu=20$ (all upper limits at 95\% confidence) \cite{cygx1_highfreq}. This  suggests that none of the above processes leads to a significant variability and, in particular, no resonances are present at their characteristic time scales.  The nature of the high frequency breaks is unclear. It may be plausible to associate them with the properties  of the emission mechanism of the hard component, for example with the details and time scales of heating and  cooling  of electrons in the  Comptonization region. Obviously these properties are different for the hard component in the soft and hard state as indicated by the difference in the high frequency power density spectra shown in the right panel in Fig. \ref{fig:pds_states}.

\subsection{(Lack of) Variability in the disk emission}
\label{sec:var_soft}

The energy spectra of black holes in the soft state often have two spectral components, a soft blackbody-like component emitted by the optically thick accretion disk and a hard power law component of thermal or non-thermal origin formed in the optically thin media  (Sect. \ref{sec:states}, Figs. \ref{fig:states_spe},\ref{fig:cygx1_bband}). The study of the variability properties of these two components in black hole X-ray novae in the early 90's with the Ginga and GRANAT\index{GRANAT} observatories  led to the conclusion that most of the variability of the X-ray emission in the soft state is usually  associated with the hard spectral component (e.g. \cite{musca_ginga, musca, kemer}), the thermal disk emission generally being much more stable. This conclusion has been later confirmed by a more rigorous and quantitative analysis \cite{churazov01,freqres_cygx1,ns_bl} whose main results are summarized below.

\begin{figure}
\centerline{
\includegraphics[width=0.8\hsize]{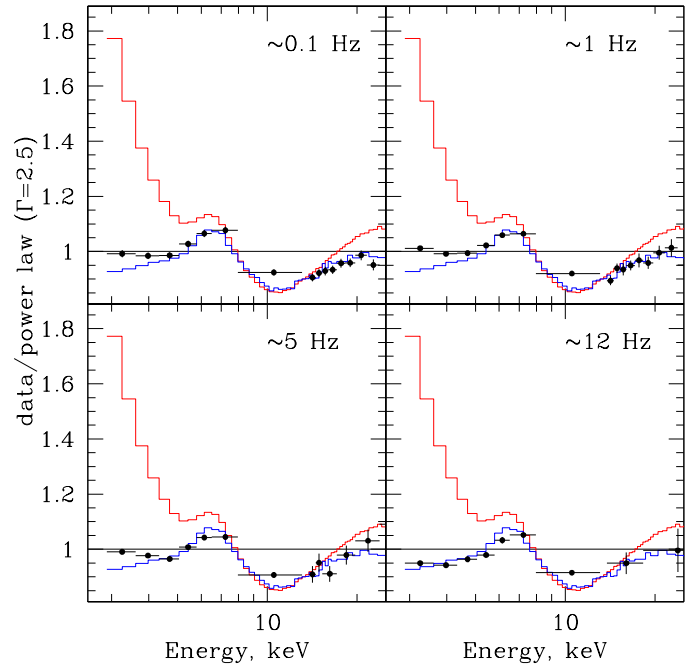}}
\caption{Frequency resolved spectra of Cyg~X-1\index{Cyg~X-1} in the soft state (June 16--18, 1996). The numbers in the upper-right corner of each panel indicate the median frequency. In each panel, the upper histogram shows the average spectrum, the lower histogram shows the frequency resolved spectrum at low frequencies, 0.002-0.033 Hz (the same data as in the left panel of Fig. \ref{fig:softcomp_var}). The spectra are plotted as ratio to a power law spectrum with photon index $\Gamma=2.5$ and low energy absorption $N_H=6\cdot 10^{21}~{\rm cm}^{-2}$.
\label{fig:freqres_soft_comp}
}
\end{figure}

The energy dependence of the variability can be characterized by the frequency-resolved energy spectrum \cite{freqres}. It is defined as  a set of Fourier-amplitudes computed from light curves in different energy channels. The Fourier-amplitudes are  integrated over the frequency range of interest and expressed in the units of flux.  Its advantages over simple energy-dependent fractional rms is the possibility to use conventional (i.e. response folded) spectral approximations and to compare its shape with shape of various spectral components present in the average energy spectrum of the source. However it can not be always regarded as the energy spectrum of the variable component. For this to be possible, certain conditions must be fulfilled \cite{ns_bl}, for example (i) independence of the shape of the frequency resolved spectrum on the Fourier-frequency  and (ii) absence of time lags between flux variations at different energies. If these two conditions are satisfied, the spectral variability  can be represented as $F(E,t)=S_0(E)+I(t)S(E)$. 

\begin{figure}[htbp]
\centerline{
\includegraphics[width=0.52\hsize]{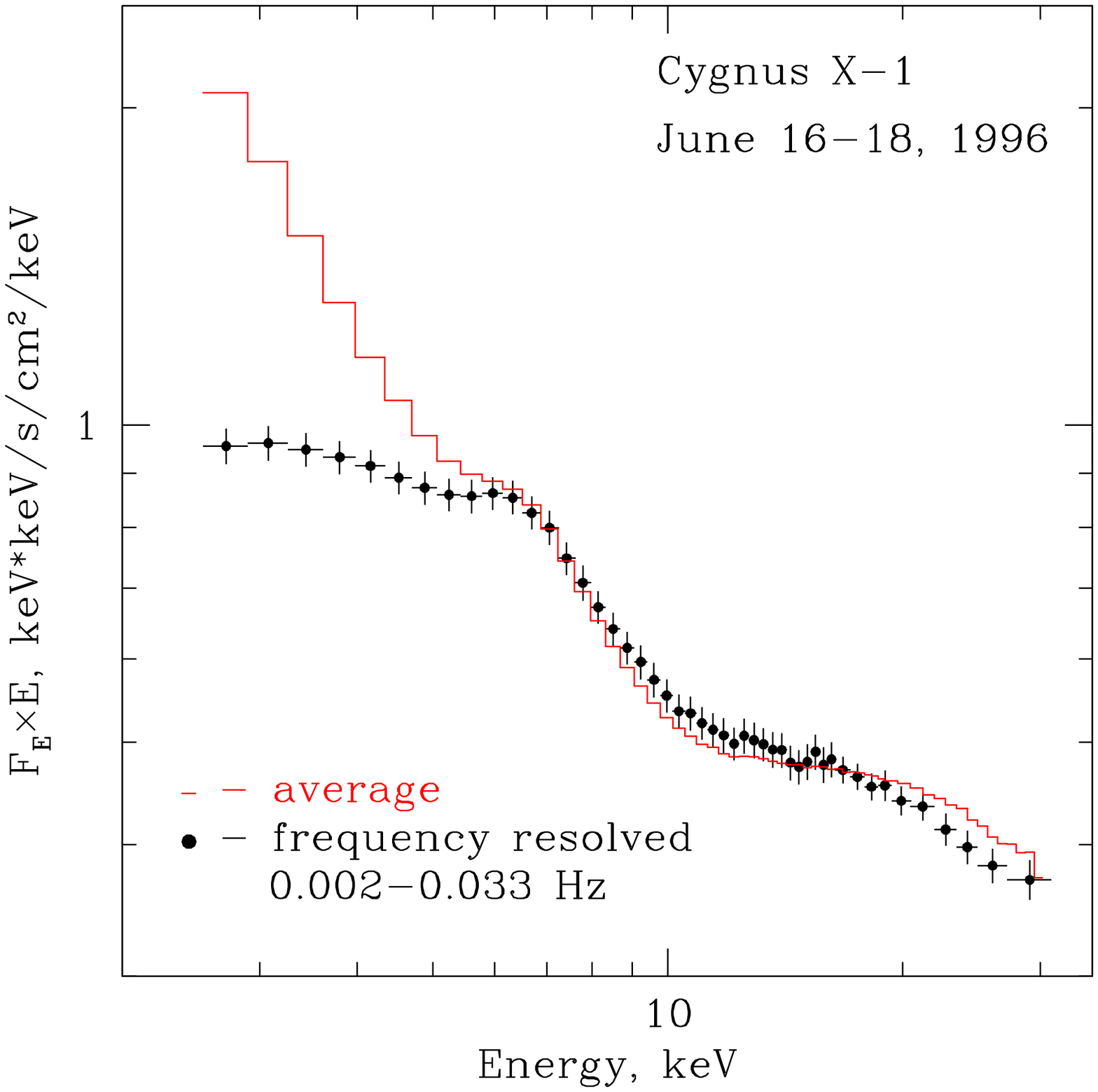}
\includegraphics[width=0.5\hsize]{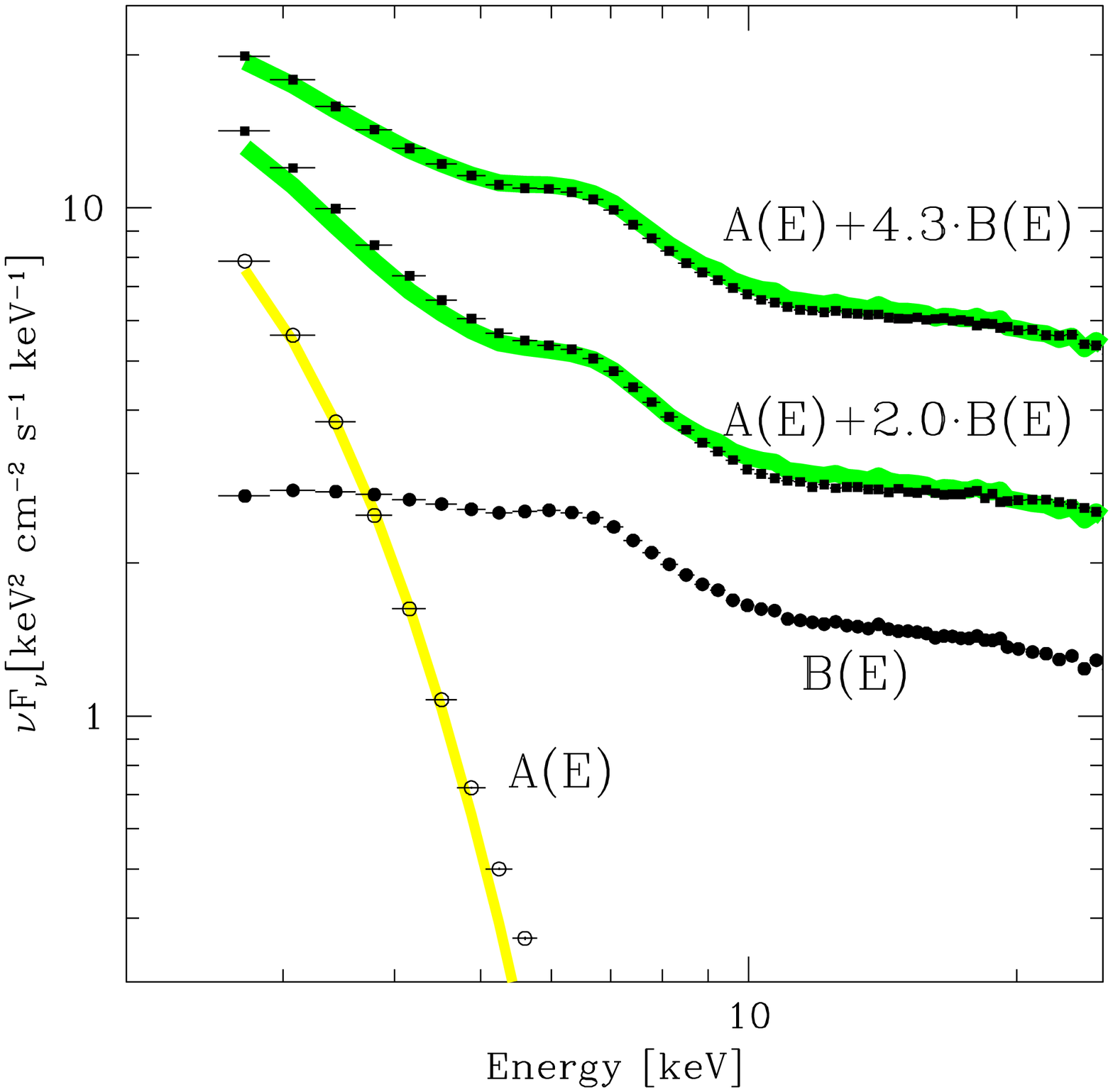}}
\caption{{\em Left:} Average and frequency-resolved spectrum of Cyg~X-1\index{Cyg~X-1}
in the 1996 soft state.
{\em Right:}
Spectra of ``constant'' (open circles -- $A(E)$) and
``variable'' (solid circles -- $B(E)$) components, derived from the
linear fits of the correlation between count rate in different
channels. The normalization of the ``variable'' component $B(E)$ is
arbitrary. For comparison, the 
light grey (yellow in the color version) curve shows the spectrum of a multicolor disk
emission with a characteristic temperature of 0.5 keV. The two upper
spectra (solid squares) were averaged over the periods of time when
the  count rate above 9 keV was high and low respectively. The dark
grey (green) lines show that these spectra can be reasonably well (within
10--15\%) approximated by a model $M(E)=A(E)+I*B(E)$ consisting of
the stable and variable spectral components  where $I$ (the
normalization of the variable component) is the only free parameter.
The right panel is adopted from \cite{churazov01}.
\label{fig:softcomp_var}
}
\end{figure}

These conditions are fulfilled in the soft state of Cyg~X-1\index{Cyg~X-1} (Fig. \ref{fig:freqres_soft_comp}): therefore the  Fourier-frequency resolved spectrum represents the energy spectrum of the variable part of  the X-ray emission. Remarkably, it coincides with the average source spectrum at $E\ge 7-8$ keV where the contribution of the disk emission becomes small (Fig. \ref{fig:softcomp_var}).  Moreover,   source light curves in different energy channels  allow a linear decomposition in the form $F(E,t)=A(E)+I(t)B(E)$ \cite{churazov01}. The constant part of the source emission $A(E)$ coincides with the spectrum of the accretion disk, while the spectrum of the variable part $B(E)$ coincides with the frequency-resolved spectrum and can be described by the  Comptonization model (Fig. \ref{fig:softcomp_var}). Thus, on time scales from $\sim 100$ msec to $\sim 500$ sec (at least) the source variability is due to variations in the flux of the hard component, whose shape is kept constant in the course of these variations, the amplitude of variability of the disk emission being significantly smaller.

This can be explained in the model of inward propagating fluctuations outlined in Sect.  \ref{sec:fluc}. The hard component is associated with the  optically thin hot flow whose aspect ratio must be comparable to unity, $h/r\sim 0.1-1$. Therefore the viscous time scale in the hot flow is $\sim 10^2-10^4$ times shorter than in the geometrically thin disk from which the soft component originates (cf. Sect. \ref{sec:vlf_break}). 
This makes the geometrically thick flow  more ``transparent" for the high frequency perturbations. Depending on the particular value of $h/r$ and of the $\alpha$-parameter, the viscosity or $\dot{M}$ fluctuations on the thermal and even dynamical time scales will be propagated inward without significant damping and will modulate the accretion rate at all smaller radii, including the region where X-ray emission is formed, leading to a significant modulation of X-ray flux.
This is not the case for the geometrically thin disk where the viscous time scale is $\sim 10^3-10^4$ times longer than the dynamical and thermal timescales; therefore high frequency perturbations will be damped (Sect. \ref{sec:fluc}). It is plausible to expect that perturbations on the thermal and dynamical time scales have larger amplitude than perturbations on the viscous time scales, thus explaining the significantly smaller variability amplitude of the disk emission. 

The lack of variability of the disk emission and its interpretation are consistent with the conclusion of the Sect. \ref{sec:vlf_break} made from completely independent  arguments. Namely, based on the location of the low-frequency break in the power density spectra of neutron stars we concluded that bulk of variability seen at low frequencies originates in the optically thin coronal flow with an aspect ratio of $h/r\sim 0.1$, rather than in the underlying geometrically thin disk. 

Finally, it should be noted that variations of the soft component may also arise from the variations of the disk truncation radius. Such variations may be absent if the geometrically thin disk extends to the last marginally stable orbit ($3r_g$ for a Schwarzschild black hole) but appear when the disk truncation radius is larger. Indeed, detailed study of the 1996 soft state of Cyg~X-1\index{Cyg~X-1} used as an example here demonstrated that on a number of occasions, mostly  at the beginning and in the end of soft state episodes,  the soft component was strongly variable. Notably, the power density spectrum during these periods had a complex shape, significantly different form the simple power law shown in Figs.\ref{fig:pds_broadband},\ref{fig:pds_states}.

\section{Variability of the reflected emission}
\label{sec:var_refl}

\begin{figure}  
\centering
\hbox{
\includegraphics[width=0.45\hsize]{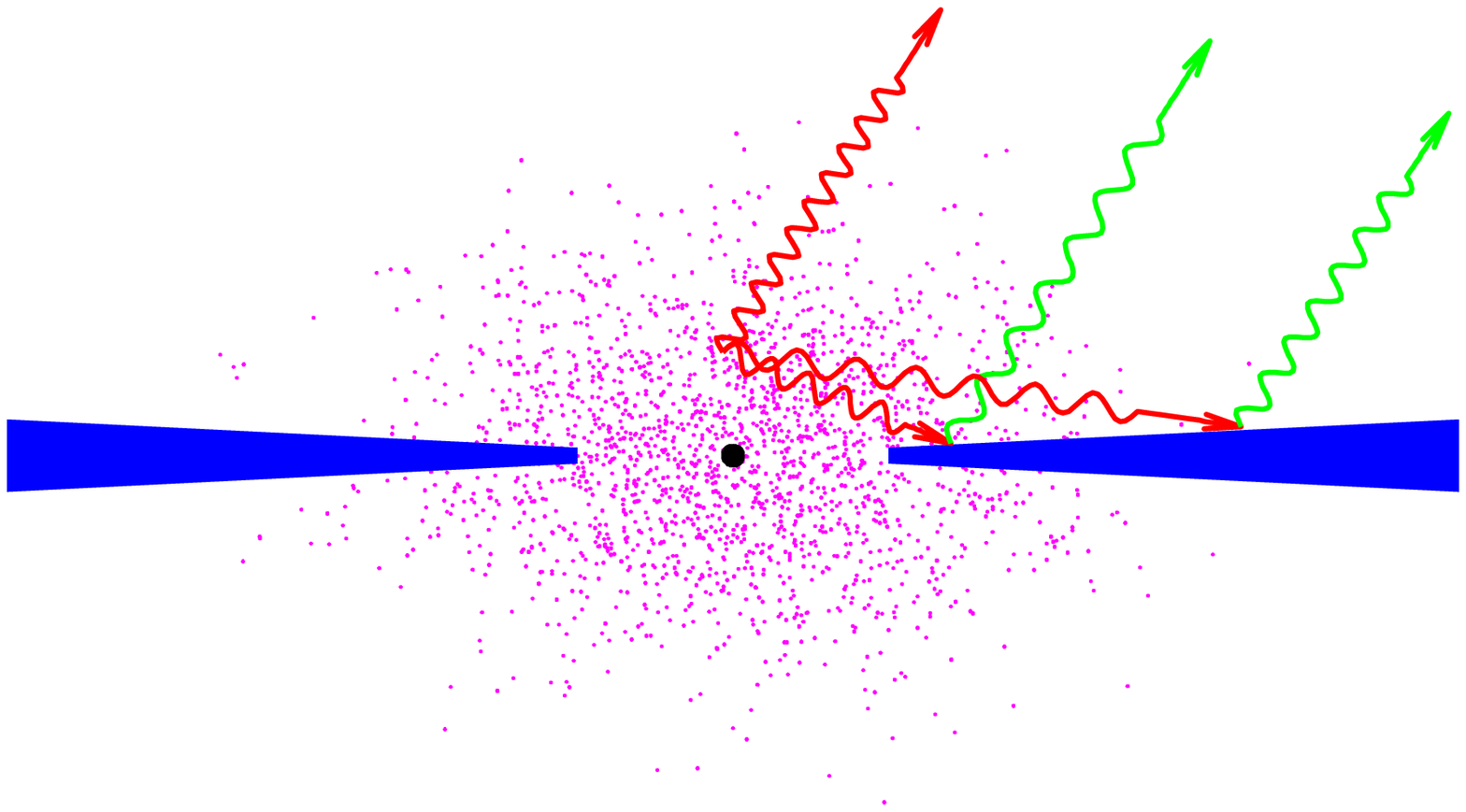}
\hspace{0.7cm}
\includegraphics[width=0.47\hsize]{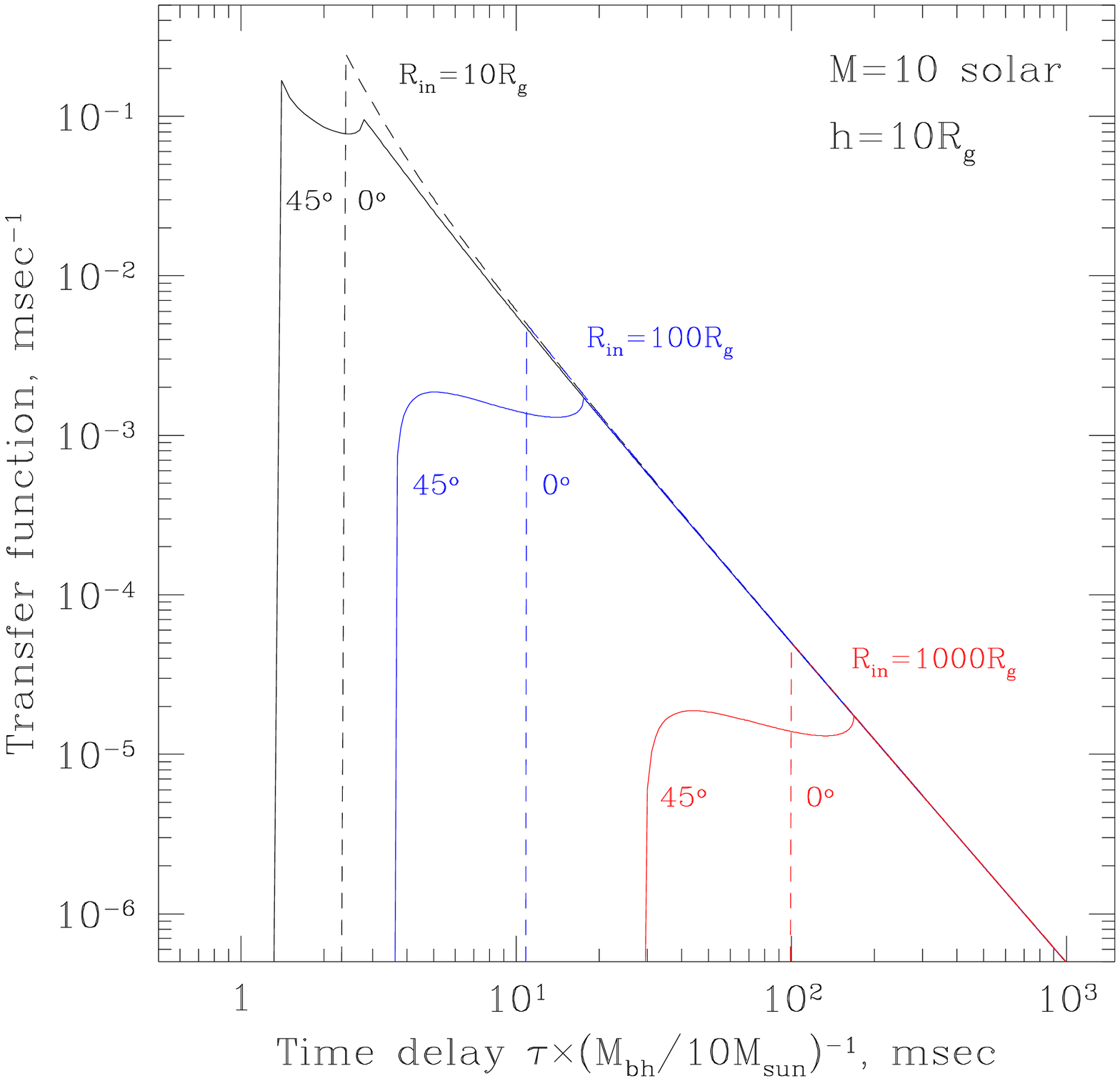}
}
\caption{The two effects defining the response of the reflected emission to variations of the Comptonized emission are the finite light travel time $\tau_d$ from the hot inner flow to the accretion disk and the finite size of the accretion disk itself, $\Delta r_d/c\sim\Delta\tau_d$ (left panel). While the first factor introduces a time delay between variations of the reflected and Comptonized components, the finite size of the disk itself leads to the suppression of high frequency variations in the reflected emission. The right panel shows the Green function of the time response of the geometrically thin disk around a $10M_\odot$ black hole (from \cite{freqres_cygx1}). The numbers at the curves mark the disk truncation radius and inclination.}
\label{fig:refl_geometry}
\end{figure}

The reflected component (Sect. \ref{sec:refl}) arises from reprocessing of the Comptonized emission in the accretion disk (Fig. \ref{fig:geometry}), therefore it should be expected to show some degree of variability, as the Comptonized radiation is strongly variable. The characteristic times of absorption/emission processes in the accretion disk are negligibly small, therefore  the main factors defining  the response of the reflected flux  to variations of the Comptonized radiation are related to the light travel times (Fig. \ref{fig:refl_geometry}). Namely, they are: (i) the finite light travel time from the source of primary radiation to the reflector,  i.e. from the Comptonization region to the accretion disk and (ii) the finite size of the reflector.  The first will introduce a time delay between variations of the reflected and Comptonized components. The amplitude of this delay is $\tau_d\sim r_d/c\sim 10 \left(r_d/100r_g\right)\left( M/10M_\odot\right)$ msec. The finite size of the disk itself will lead to suppression of the high frequency variations in the reflected component -- the accretion disk acts as a low-pass filter. It seems to be possible to estimate the cut-off frequency  from the size of the accretion disk making the main contribution to the reflected flux, $\Delta r_d\sim r_d$, thus leading to $f_{cut}\sim \Delta\tau_d^{-1}\sim \left(\Delta r_d/c\right)^{-1}\sim 100$ Hz, i.e. beyond the frequency range of the bulk of observed variability (Fig. \ref{fig:pds_states}). However, calculations of the transfer function \cite{freqres_cygx1} show that at the frequency $f_{cut}\sim \Delta\tau_d^{-1}$ the variability signal is suppressed by a significant factor, of $\sim 10-20$, whereas a noticeable suppression of variability, by a factor of $\sim 2$ or more occurs at frequencies $\sim 10$ times lower. 
As these effects directly depend on the mutual location and geometry of the Comptonization region and accretion disk, their observation is a powerful tool in studying the geometry of the accretion flow.

\begin{figure}
\centering
\includegraphics[width=0.475\hsize]{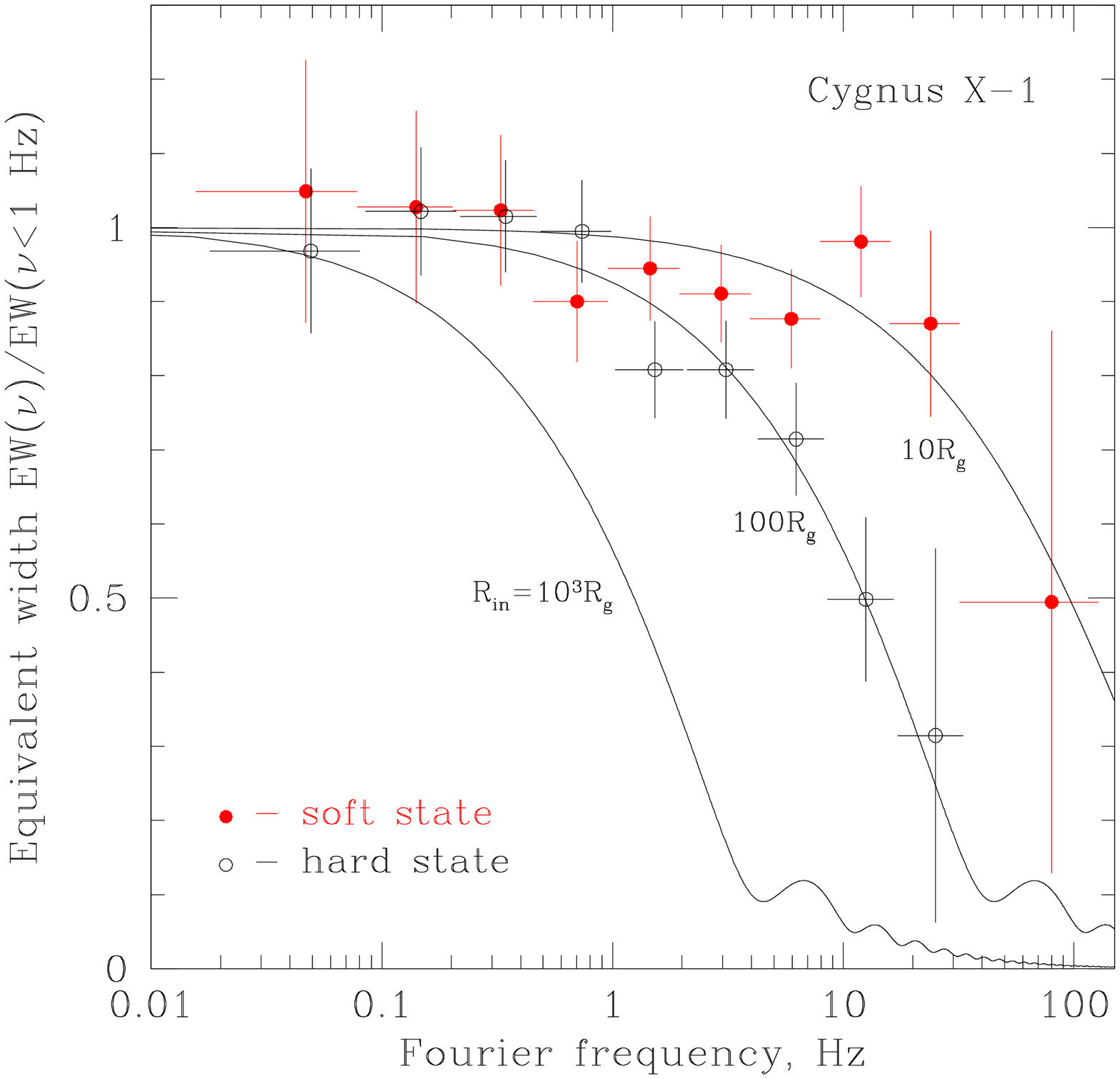}
\includegraphics[width=0.51\hsize]{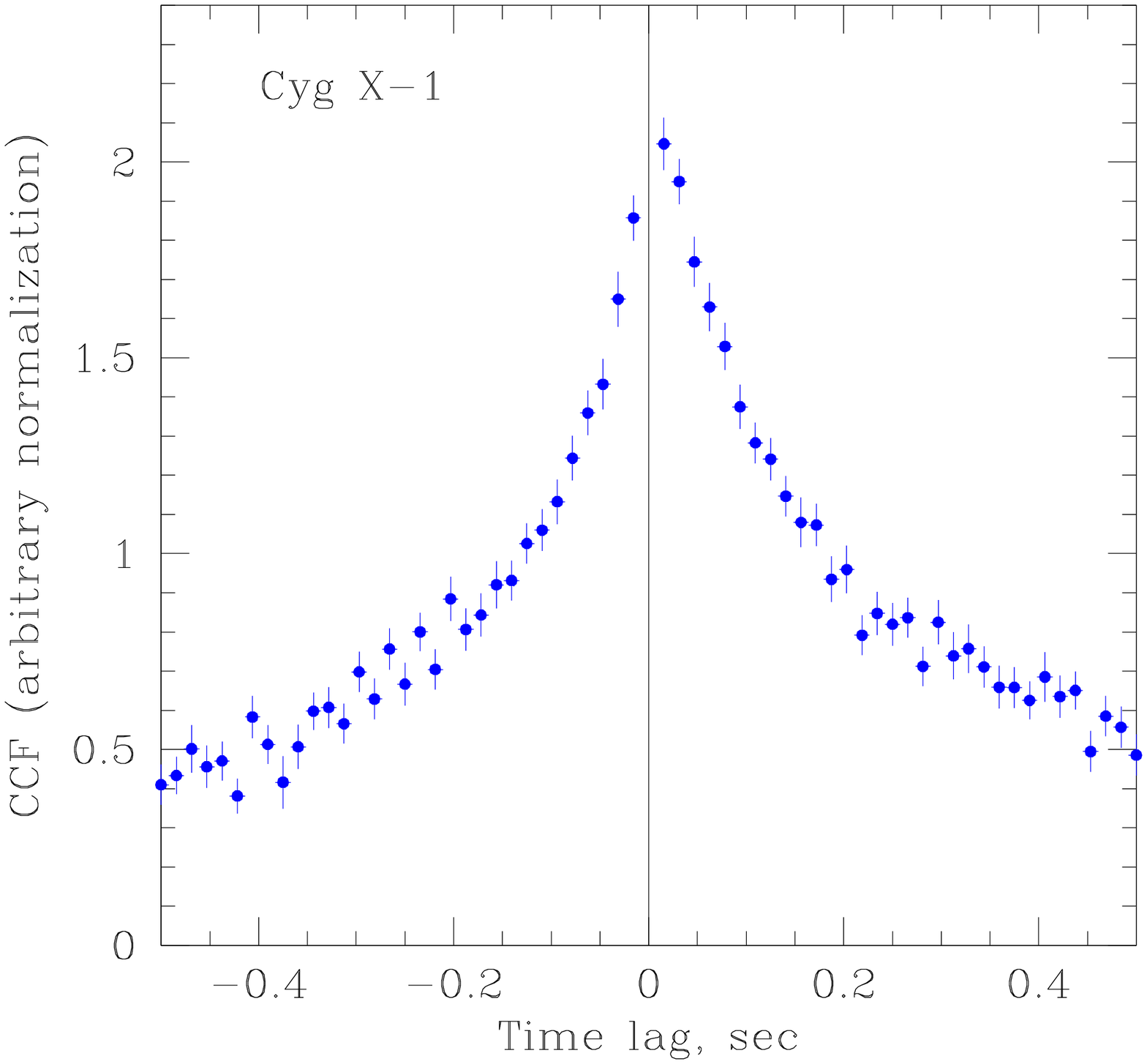}
\caption{Suppression of high frequency variations (left) and time delay (right) in the reflected emission.
The left panel shows the equivalent width of the iron fluorescent line (= ratio of the fractional rms of the iron line flux to that of the underlying continuum) vs. Fourier frequency. The model curves are  for an isotropic point source at the height $h=10r_g$ on the axis of a flat disk with inner truncation  radius of 
$10,100$ and $1000 r_g$ (assuming a $10M_{\odot}$ black hole) and with
inclination angle of $50$ degrees. A narrow line was assumed in calculations. The right panel shows the cross-correlation function of the iron line emission and the power law continuum. Positive lags mean delay of the line emission with respect to the continuum.}
\label{fig:freqres-refl}
\end{figure}

As suggested in \cite{freqres, freqres_cygx1}, the variability of the reflected emission can be studied using the methods of frequency-resolved spectroscopy (Sect. \ref{sec:var_soft}). This method is based on the analysis of Fourier amplitudes and therefore washes out phase information, precluding  the study of time delay effects. However, it provides a convenient way to explore the frequency dependence of the variability of the iron line flux  and to compare it with that of the continuum emission which is dominated by the Comptonized radiation.  Using this method, \cite{freqres, freqres_cygx1} showed that  in the soft spectral state, variations of the reflected component have the same frequency dependence of the rms amplitude as the Comptonized  emission up to frequencies $\sim 30$ Hz (Fig. \ref{fig:freqres-refl}). This would be expected if, for instance, the reflected flux was reproducing, with a flat response, the variations of the primary radiation down to time-scales of $\sim 30-50$ ms. The sensitivity of their analysis was insufficient to study shorter time-scales.  In the hard spectral state, on the contrary, the variability of the reflected flux is significantly suppressed in comparison with the direct emission on time-scales shorter than $\sim 0.5-1$ s.  These findings are to be compared with the predictions of the simple model for the time response of the disk to variations of the primary emission.
Assuming that suppression of the short-term variability of the reflected emission is caused by the finite light-crossing time of the accretion disk, one can estimate the truncation radius of the accretion disc, $r_d\sim 100 r_g$ in the hard spectral state and $r_d \le 10r_g$ in the soft spectral state (Fig. \ref{fig:freqres-refl}). This agrees well with the interpretation of the spectral states\index{states} in black holes in the ``sombrero" geometry of the accretion flow (Sect.  \ref{sec:states}).

In order to study the time delay in the reflected emission, one would need to separate  the reflected component from the main emission in time-resolved energy spectra. This can be done most easily for the fluorescent iron line, while the reflection continuum is more difficult to separate due to its breadth in the energy domain. An attempt to perform such an analysis based on RXTE/PCA data of Cyg~X-1\index{Cyg~X-1} in the hard state is presented in the right panel in Fig. \ref{fig:freqres-refl}. In each bin of the light curve with 16 msec resolution, the spectrum in the 3--20 keV band  was linearly decomposed in to power law component, reflected continuum\index{Compton reflection} and 6.4 keV iron line emission. Shown in Fig. \ref{fig:freqres-refl} is the cross-correlation of the light curves of the iron line flux and the power law component. The cross-correlation function shows a clear asymmetry, suggesting a time delay of the order of $\sim 10-15$ msec. The amplitude of the possible time delay corresponds to a 
disk truncation radius of $\sim 100 r_g$, in good agreement with the number obtained from the analysis of high-frequency variations in the iron line flux (left panel in Fig. \ref{fig:freqres-refl}) and also with the numbers tentatively suggested by spectral analysis.

The results shown in Fig. \ref{fig:freqres-refl} seem to suggest a rather consistent picture and to favor to the ``sombrero"-type configuration of the accretion flow, with the spectral state transition being related to a change of the disk truncation radius. However, a caveat is in order. While the rms amplitude behavior of the  iron line emission is a rather robust observational result, the search of the time delay of the reflected component requires  separation of  line and continuum emission components. On $\sim 10$ msec time scales, this can not be done through direct spectral fitting because of  insufficient statistics, even with the large collecting area of the PCA instrument aboard RXTE, and requires more sophisticated data analysis techniques, for example the one used to produce the cross-correlation function shown in Fig. \ref{fig:freqres-refl}.  Secondly, although the interpretation of these results in terms of finite light travel times is the most simple and straightforward, alternative scenarios are also possible as discussed in length in \cite{freqres_cygx1}. Nevertheless, the former interpretation  is, in my view, the simplest and most attractive one and is further supported by the results of the spectral analysis, as described in the next section.

\section{$R-\Gamma$ and other correlations}
\label{sec:correl}

\begin{figure}
\centering
\includegraphics[width=0.6\hsize]{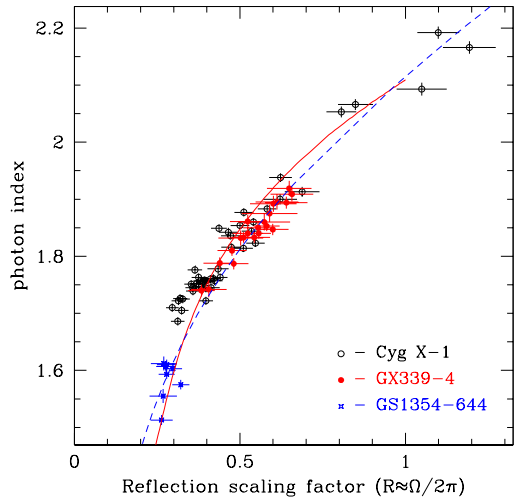}
\caption{The $R-\Gamma$ correlations between the photon index $\Gamma$ of the Comptonized
radiation and the relative amplitude of the reflected component $R$ in three
``well-behaved'' black hole systems Cyg~X-1\index{Cyg~X-1}, GX~339-4\index{GX~339-4} and
GS~1354-644\index{GS~1354-644}. 
The solid and dashed lines show the the dependence $\Gamma(R)$
expected in the disk-spheroid and in the plasma ejection models
discussed in the text. Adopted from \cite{urumuqi}
\label{fig:r-gamma}
}
\end{figure}

\begin{figure}
\centering
\includegraphics[width=0.6\textwidth]{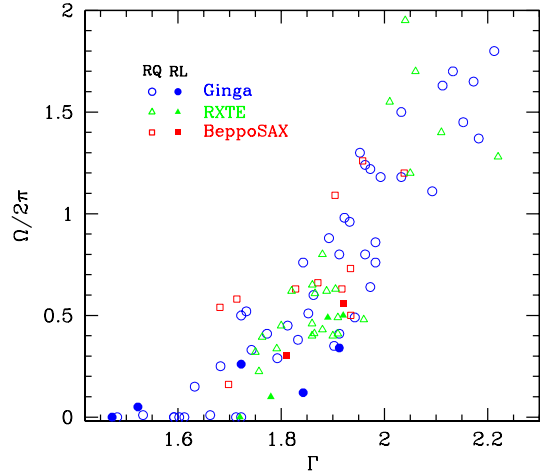}
\caption{$R-\Gamma$ relation for supermassive black holes, from \cite{aaz2003}}
\label{fig:r-gamma-agn}
\end{figure}

Observations show that  spectral and timing parameters of accreting black holes often change in a correlated way, e.g. \cite{paper1,gx339,cygx1_correl}. One of  the most significant correlations is the one between the photon index  of the Comptonized spectrum, the amplitude of the reflected component and the characteristic frequencies\index{characteristic frequencies} of aperiodic variability (Figs. \ref{fig:r-gamma},\ref{fig:refl-qpo}). The correlation between spectral slope and reflection amplitude is also known as $R-\Gamma$ correlation \cite{paper1,zdz1}. Its importance is further amplified by the fact that it is also valid for supermassive black holes (Fig. \ref{fig:r-gamma-agn})  \cite{aaz2003}.

\subsection{$R-\Gamma$ correlation}
\label{sec:r-g}

\begin{figure}
\centering
\includegraphics[width=0.6\textwidth]{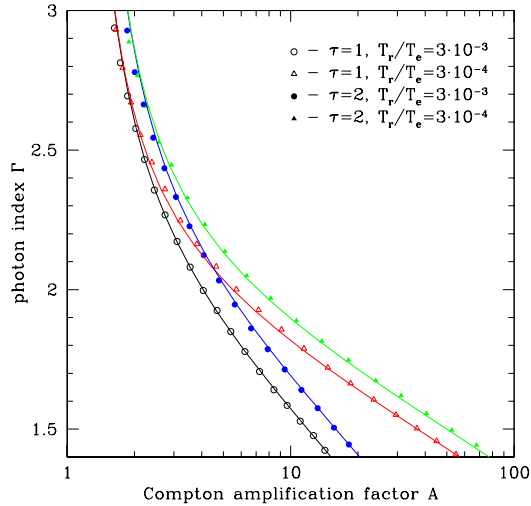}
\caption{Relation between the photon index of Comptonized radiation $\Gamma$ and the Compton amplification factor $A$.  The symbols show results of Monte-Carlo simulations assuming spherical geometry for different values of parameters of the Comptonization region and soft seed photons. The solid lines are calculated using eq.(\ref{eq:g-a}). Adopted from \cite{urumuqi}}
\label{fig:g-a}
\end{figure}

The spectrum formed by the unsaturated Comptonization of low frequency seed photons with characteristic temperature $T_{bb}$ on hot electrons with temperature $T_e$ has a nearly power law shape in the energy range from $\sim 3 kT_{bb}$ to $\sim kT_e$ \cite{st80}. For the parameters typical for black hole X-ray binaries in the hard
spectral state, this corresponds to the energy range from $\sim 0.5-1$ keV to $\sim
50-100$ keV (e.g. Fig. \ref{fig:geometry}). The photon index $\Gamma$ of the Comptonized spectrum depends in a
rather complicated way on the parameters of the Comptonizing media, primarily on the electron temperature and the Thompson optical depth  \cite{st80}. It is more meaningful to relate $\Gamma$ to the Comptonization parameter $y$ or, nearly equivalently, to the Compton amplification factor $A$. The latter describes the energy balance in the corona and is defined as the ratio of the energy deposition rate into hot electrons and the energy flux brought into the
Comptonization region by soft seed photons. The concrete shape of the $\Gamma(A)$  relation
depends on the ratio $T_{bb}/T_e$ of the temperatures of the seed
photons and the electrons, the Thomson optical depth and the
geometry, but broadly speaking, the higher the Compton amplification
factor, the harder is the Comptonized spectrum
\cite{st89,enh,haardt,kemer}. This is illustrated by the results of
Monte-Carlo simulations shown in Fig. \ref{fig:g-a}.

The strength of the reflected component in the
spectrum depends on the fraction of the Comptonized radiation
intercepted by the accretion disk  (Sect. \ref{sec:refl}). The latter is defined by the
geometry of the accretion flow, namely, by the solid angle
$\Omega_{disk}$ subtended by the accretion disk as seen from the
corona.  In addition, the spectrum of the reflected emission depends on the ionization state of the disk, 
in particular its low energy part which is formed by the interplay  between  Thomson scattering and  photoabsorption and fluorescence by metals. The problem is further complicated by the fact that the ionization state of the disk can be modified by the Comptonized radiation. 

Observations show that there is a clear correlation between the photon index of the Comptonized radiation $\Gamma$ (i.e. the Comptonization parameter)  and the relative amplitude of the reflected component $R$ (Fig. \ref{fig:r-gamma}). Softer spectra (lower value of  the Comptonization parameter $y$ and of the Compton amplification factor $A$) have stronger reflected component, revealing itself, for example, via a larger equivalent width of the iron fluorescent line.
The existence of this correlation suggests that there is a positive correlation between the fraction of the Comptonized radiation intercepted by the accretion disk and the energy flux of the soft seed photons to the Comptonization region \cite{zdz1}.  This is a strong argument in favor of the accretion disk  being the primary source of  soft seed photons to the Comptonization region. Indeed, in the absence of strong beaming effects a correlation between $\Omega_{disk}$ and the seed photons flux should be expected since an increase of the solid angle of the disk seen by the hot electrons ($=\Omega_{disk}$) should generally lead to the increase of the fraction of the disk emission reaching the  Comptonization region.

\subsection{Toy models}
\label{sec:toy_models}

We illustrate the above considerations with two simple and  idealized 
models having a different cause of change of  $\Omega_{disk}$.  
In the first, the disk--spheroid model (cf. ``sombrero" configuration, Sect.  \ref{sec:geometry}), an optically-thin
uniform hot sphere with radius $r_{sph}$, the source of the hard Comptonized radiation is surrounded by an
optically-thick cold disk with an inner radius $r_{disk}$, $\Omega_{disk}$ depending on the ratio
$r_{disk}/r_{sph}$. 
Propagation of the the disk towards/inwards the hot  
sphere (decrease of $r_{disk}/r_{sph}$) leads to an increase of the
reflection scaling factor $R$, a decrease of the the Compton amplification
factor $A$ and a steepening of the Comptonized spectrum.
In such context the model was first studied by Zdziarski,
Lubinski \& Smith \cite{zdz1}. In the second, the
plasma ejection model, the value of the $\Omega_{disk}$ is defined by the
intrinsic properties of the emitting hot plasma, particularly by its
bulk motion with mildly relativistic velocity towards or away from the
disk, which itself remains unchanged \cite{belob}.  In the case of
an infinite disk, values of the reflection scaling factor $R$ below and above unity correspond
to the hot plasma moving respectively away from and towards the disk.

Both models predict a relation between
reflection $R$ and Compton amplification factor $A$ which can be
translated to $\Gamma(R)$  given a dependence $\Gamma(A)$ of the
photon index of the Comptonized spectrum on the amplification factor.  
The  relation between $\Gamma$ and $A$ can be approximated by:
\begin{equation}
A=(1-e^{-\tau_T})\cdot\frac{1-\Gamma}{2-\Gamma}\cdot
\frac{\left({T_e}/{T_{bb}}\right)^{2-\Gamma}-1}{\left({T_e}{T_{bb}}\right)^{1-\Gamma}-1}+e^{-\tau_T}
\label{eq:g-a}
\end{equation}
This formula is based on  a representation of the Comptonized spectrum
by a power law in the energy range $kT_{bb}-kT_e$ and takes into account that a fraction $e^{-\tau_T}$ of the soft radiation will leave the Comptonization region unmodified. Despite its simplicity, it agrees with the
results of the Monte-Carlo calculations with reasonable accuracy for
optical depth $\tau_T\sim 1$ and $T_{bb}/T_e\sim 10^{-5}-10^{-3}$
(Fig. \ref{fig:g-a}).

The expected $\Gamma(R)$ relations are shown in 
Fig. \ref{fig:r-gamma}. With a proper tuning of the parameters, both
models can reproduce the observed shape of the $\Gamma(R)$ dependence
and in this respect are virtually indistinguishable. 
The models plotted in Fig. \ref{fig:r-gamma} were calculated with  the
following parameters: the disk--spheroid model assumes the disk 
albedo $a=0.1$, Thomson optical depth of the cloud $\tau_T=1$ and the
ratio of the temperature of the seed photons to the electron
temperature $T_{bb}/T_e=10^{-4}$. I note that the latter value is too small for stellar mass black holes, a more realistic one being in the range $T_{bb}/T_e=10^{-3}$. However, this  does not invalidate the model as a number of significant effect are ignored in this calculation which can modify the $\Gamma(R)$ dependence, for example disk inclination, gravitational energy release in the disk etc. The exact importance of these factors is  yet to be determined. 
The plasma ejection model parameters are:
$a=0.15$, $\tau_T=1$, $T_{bb}/T_e=3\cdot 10^{-3}$ and $\mu_s=0.3$.
The observed range of the reflection $R\sim 0.3-1$ and the slope
$\Gamma\sim 1.5-2.2$ can be explained assuming variation of the disk
radius from $r_{disk}\sim r_{sph}$ to $r_{disk}\sim 0$ in the
disk-spheroid model or variation of the bulk motion velocity from
$v\sim 0.4 c$ away from the disk to $v\sim 0$ in the plasma ejection model.  

Of course these models are very simple and schematic and the real configuration of the accretion flow  is likely to be far more complex. They are presented here with the only purpose to demonstrate that simple geometrical considerations can successfully explain the observed correlation between parameters of the Comptonized and reflected emission in black holes. More sophisticated scenarios are considered for
example in \cite{dyn_corona}.

\subsection{Characteristic frequencies of variability}

\begin{figure}
\centering
\includegraphics[width=0.6\hsize]{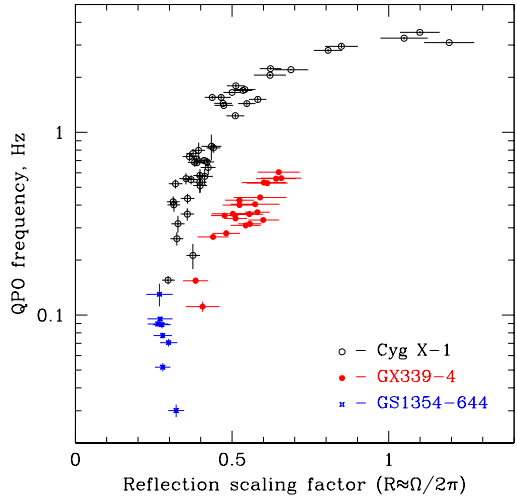}
\caption{The correlations between the amplitude of the reflected
component $R$ and characteristic frequencies\index{characteristic frequencies} of aperiodic variability. The frequency of the second peak in the $\nu\,P_\nu$ plot (Fig.\ref{fig:pds_states}) was used to represent the latter. 
\label{fig:refl-qpo}
}
\end{figure}

As discussed in the Sect. \ref{sec:high_freq}, the power density spectra of accreting black holes above $f\ge 10^{-2}$ Hz  have a number of bumps and peaks which define several frequencies characterizing the variability time scales in the accretion flow.
These frequencies  usually correlate with each other therefore almost any of them may be used to represent characteristic variability timescales. As illustrated by Fig. \ref{fig:refl-qpo}, a tight correlation exists between the reflection amplitude $R$ and the characteristic variability frequencies -- an increase of the amplitude of the reflected component in the energy spectrum is accompanied by an increase of the variability frequencies. This correlation covers a remarkably broad dynamical range, nearly two orders of magnitude in frequency.

Although the precise nature of the characteristic noise
frequencies is still unknown, it is plausible that they are associated
with the Keplerian and viscous time scales of the disk and corona
at various characteristic radii, for example, at the
truncation  radius of the disk (Sect. \ref{sec:high_freq}). 
If this is true, the correlation between $R$ and $\nu$ can be easily understood, at least qualitatively.
Indeed, in the truncated disk
picture the increase of reflection is caused by the inward
propagation of the inner disk boundary, hence, it is
accompanied by an increase of the Keplerian frequency at the disk
truncation radius and corresponding increase of the characteristic
noise frequencies.

\subsection{Doppler broadening of the iron line}

\begin{figure}[htbp] 
\centerline{    
\hbox{
\includegraphics[width=0.48\hsize]{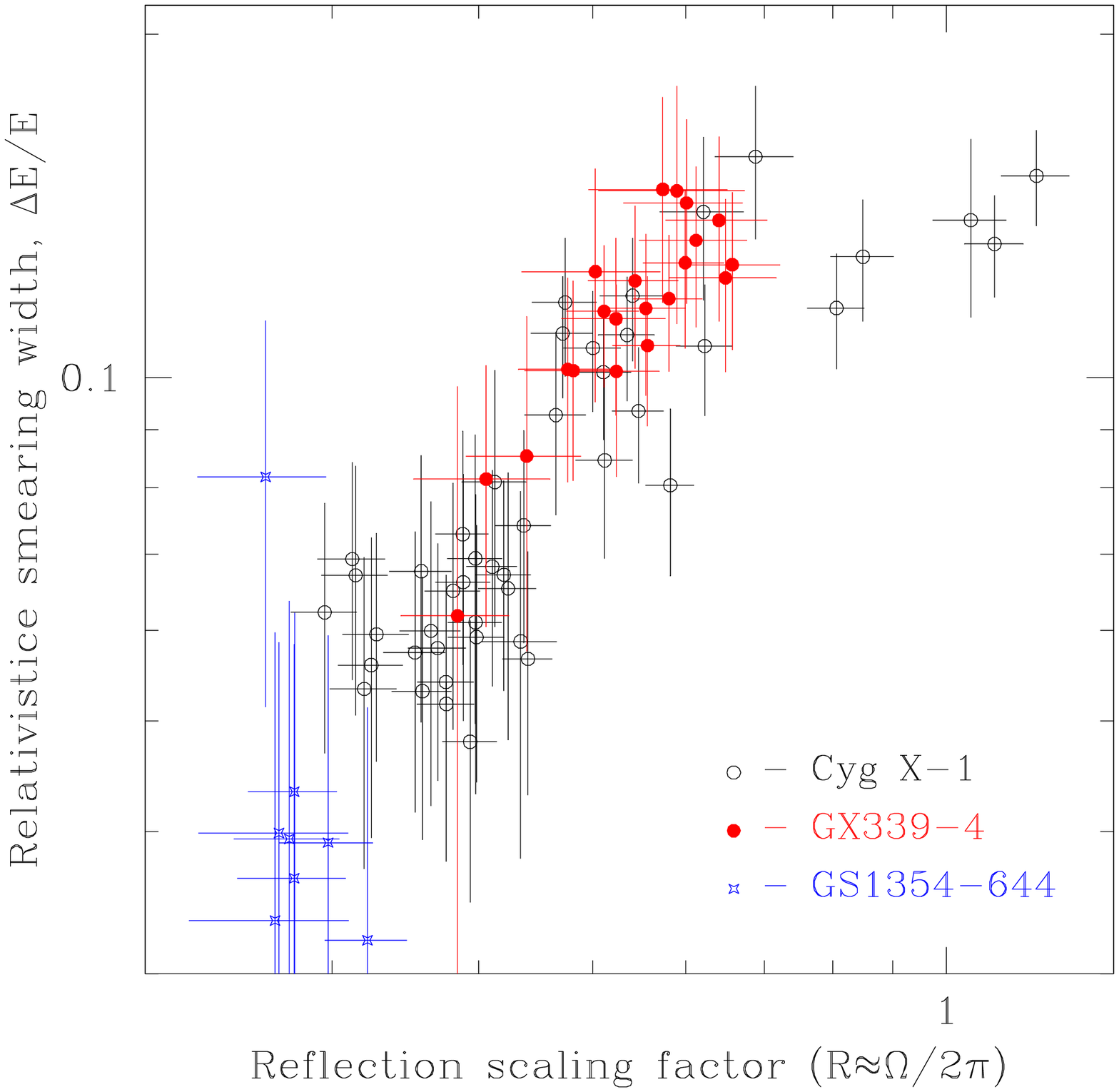}
\includegraphics[width=0.48\hsize]{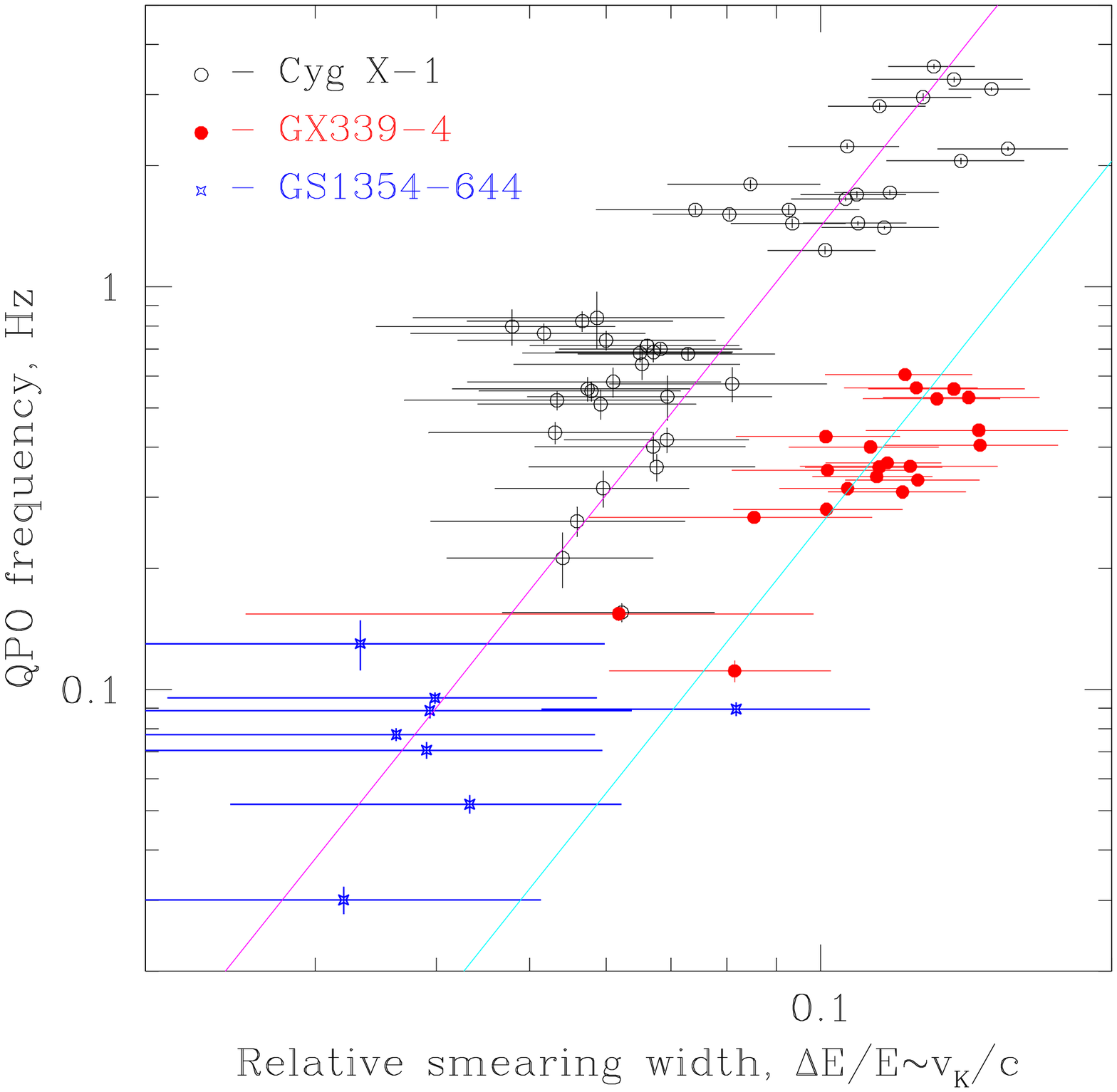}
}}
\caption{Relation between the Doppler broadening\index{Doppler broadening} of the iron
line and the reflection scaling factor ({\em left}) and the
characteristic frequency of variability ({\em right}). 
The straight lines in the right panel show the dependence $\Delta
E/E\propto \nu_{QPO}^{1/3}$ expected in the truncated disk picture if
the characteristic frequencies\index{characteristic frequencies} of variability were proportional to the  
Keplerian frequency at the inner boundary of the accretion disk.}
\label{fig:doppler}
\end{figure}

The spectrum of the emission, reflected from a Keplerian
accretion disk, is modified by special and general relativity effects
\cite{fab1}, in particular the width of the fluorescent line of iron is
affected by the Doppler effect due to Keplerian motion in the disk (Sect. \ref{sec:refl}). 
If the increase of the reflection amplitude is caused by the decrease of the inner
radius of the accretion disk, a correlation should be expected between
the amplitude of reflected emission and its Doppler broadening\index{Doppler broadening}, in
particular the Doppler width 
of the iron line. Such a correlation is a generic
prediction of  the truncated disk models and might be used to
discriminate between different geometries of the accretion flow.    
The energy resolution of RXTE/PCA, whose data has been used for this
study \cite{xrt2003}, is not entirely adequate for the task to accurately measure the relativistic smearing of
the reflection features. However, the data shown in
Fig. \ref{fig:doppler} suggests a correlated behavior of the
reflection and the Doppler broadening of the fluorescent line of iron.

Speculating further, if the characteristic frequencies of variability
are proportional to the Keplerian frequency at the inner boundary of
the disk, they should scale as 
$$
\nu_{QPO}\propto\omega_K\propto r_{ disk}^{-3/2}
$$
As the reflected emission is likely to originate primarily
from the innermost parts of the accretion disk, closest to the source
of Comptonized radiation, the effect of the Doppler broadening\index{Doppler
broadening} should be proportional to  the Keplerian linear velocity at
the inner edge of the disk:
$$
\frac{\Delta E}{E}\propto \frac{{\rm v}_K}{c}\,\sin i
\propto r_{disk}^{-1/2}\,\sin i
$$ 
Therefore, one might expect that the characteristic frequencies\index{characteristic frequencies} of
variability and the Doppler broadening\index{Doppler broadening} of the fluorescent line should
be related via:
$$
\nu_{QPO}\propto \frac{\left(\Delta E/E\right)^3}{M_{\rm BH}\; \sin^3 i}
$$ 
where $M_{\rm BH}$ is the black hole mass and $i$ is inclination of
the binary system. The PCA data indicate that such dependence
might indeed be the case (Fig. \ref{fig:doppler}, right panel). However,  an independent confirmation by observations with higher energy resolution instruments is still needed.

\bigskip

Thus, observations speak in favor of the truncated disk scenario. However this point of view is not universally accepted and alternative interpretations and scenarios are being investigated. The counter arguments are based, for example, on the possible detection of the relativistic broad Fe K line and cool disk emission component in the hard state spectra of several black holes, suggesting that the optically thick disk might be present in the vicinity of the compact object in the hard state as well.  For the detailed discussion the interested reader is referred to the original work, e.g. \cite{miller06} and references therein.

\section{Comparison with neutron star binaries}
\label{sec:ns}

Neutron star radii are most likely in the range $\sim 10-15$ km, i.e. of the order of $\sim 3-4r_g$ for a $\sim 1.4M_\odot$ object. This is comparable to the radius of the last marginally stable Keplerian orbit around a non-rotating black hole, hence  the efficiency $\eta=L_X/\dot{M}c^2$ of accretion onto a neutron star is not much different from that on a black hole.  Therefore at comparable $\dot{M}$ accreting black holes and neutron stars would have comparable X-ray luminosities (but see below regarding the contribution of the boundary layer). The Eddington luminosity limit, however, is proportional to the mass of the central objects, and the typical maximum luminosity is by $M_{BH}/M_{NS}\sim 5-10$ times smaller for neutron stars. This is in general agreement with observations of peak luminosities of black hole transients and their  comparison to luminosities of persistent and transient neutron star systems in the Milky Way.

\begin{figure}[htbp] 
\centerline{    
\hbox{
\includegraphics[width=0.5\hsize]{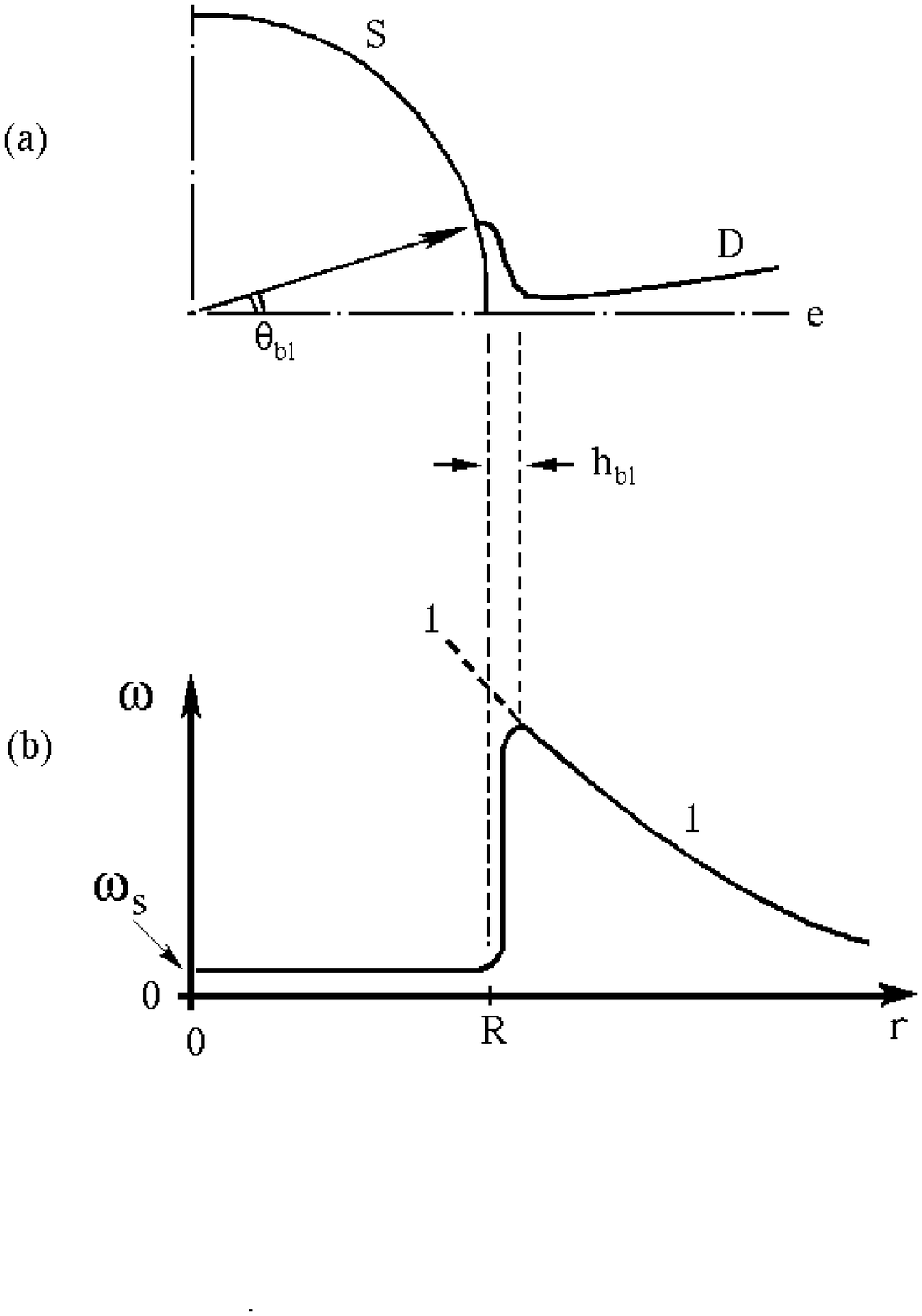}
\includegraphics[width=0.53\hsize]{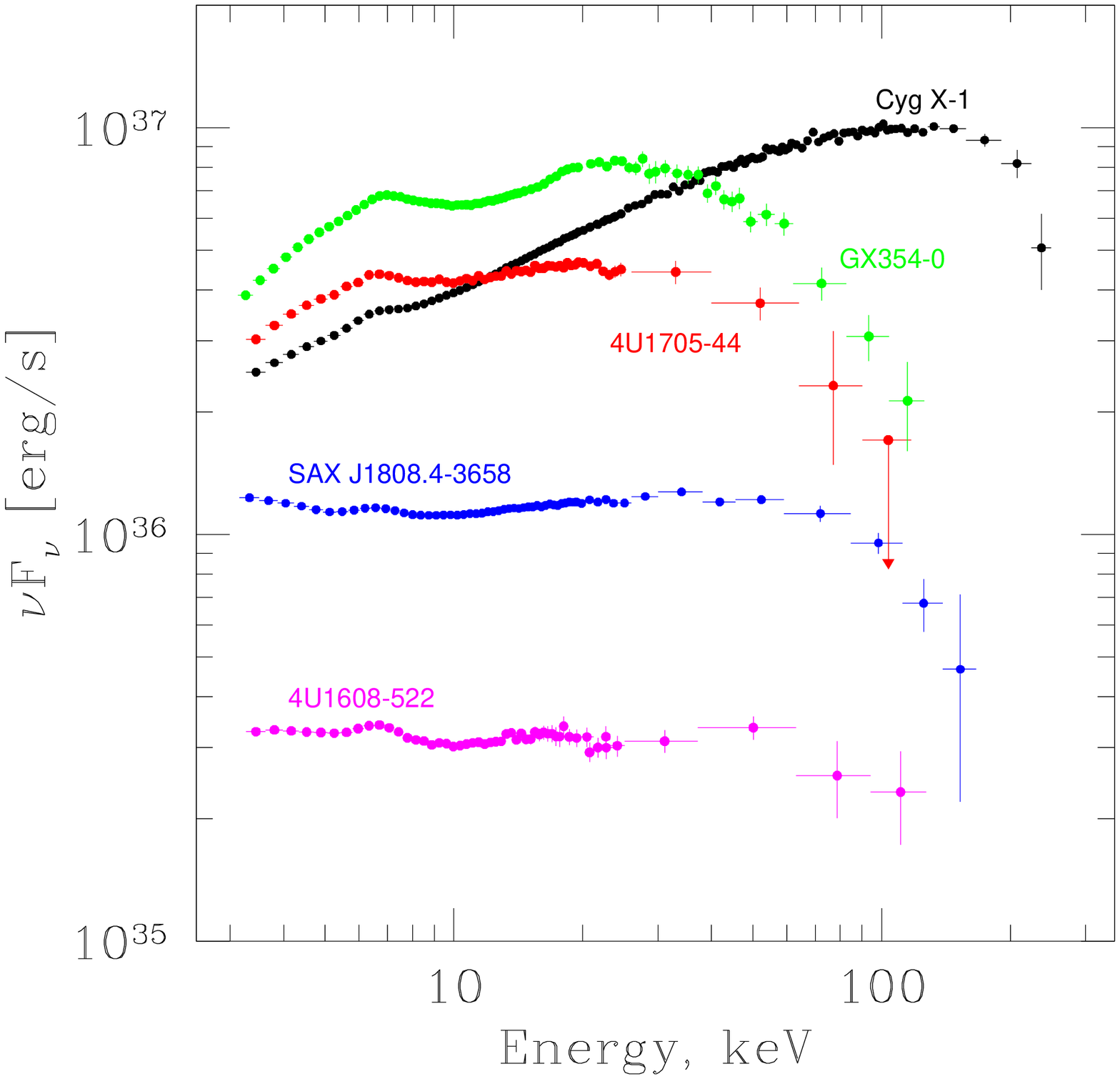}
}}
\caption{The left panels show the geometry of the spreading layer on the surface of the neutron star (upper)  and radial dependence of the angular velocity of accreting matter (lower).  Adopted from \cite{inogamov99}. On the right:  Hard state spectra of several weakly magnetized accreting neutron stars and of a black hole Cyg~X-1\index{Cyg~X-1}. Based on the  data of RXTE observations.}
\label{fig:bhns_spe}
\end{figure}

The fact that the size of a neutron star  is of the order of $\sim 3 r_g$, i.e. is comparable to the radius of the last marginally stable Keplerian orbit around a black hole,  also suggests that the structure of the accretion disk in both cases may be similar. This is indirectly confirmed by the existence of two spectral states\index{states} in accreting neutron stars, which properties are qualitatively similar to black holes (Fig. \ref{fig:states_spe}).  A more direct argument is  presented by the similarity of the accretion disk spectra in the soft state, which can be described by the same spectral models in black holes and neutron stars (e.g.\cite{ns_bl}). Moreover, the power density spectra of the accretion disk in both types of systems follow a $P_\nu\propto \nu^{-\alpha}$ power law with $\alpha$ close to $\sim 1$ (Sect. \ref{sec:variability} and \cite{ns_bl}). The strong magnetic field which may exist around young neutron stars can change the picture, causing disruption of the accretion disk at a large distance from the neutron star and modifying dramatically the structure of the accretion flow inside the magnetospheric radius $r_m\gg r_{NS}$.  This may lead to the phenomenon of X-ray pulsations common in high-mass X-ray binaries and is not  considered in this chapter.

The main qualitative difference between the two types of compact objects is obviously the existence of a solid surface of the neutron star which is absent in the case of a black hole.  Neutron star rotation frequencies are typically in the few hundred Hz range, i.e.  $\sim$ several times smaller than the Keplerian frequency  near its surface, $\nu_K\sim$ kHz. Therefore a boundary or spreading layer will appear near the surface of the neutron star where accreting matter decelerates from Keplerian rotation in the accretion disk down to the neutron star spin frequency and settles onto its surface (Fig. \ref{fig:bhns_spe}, left panels). 
In Newtonian approximation, half of the energy of the test particle on a Keplerian orbit is in the form of kinetic energy of the Keplerian rotation, hence for a non-rotating neutron star half of the energy release due to accretion  would take place in the boundary/spreading layer.
The effects of general relativity can  increase this fraction, e.g. up to $\sim 2/3$  in the case of a neutron star with radius $r_{NS}=3 r_{g}$ \cite{ss86,sibg00}. 
Rotation of the neutron star and deviations of the space-time geometry from Schwarzschild metric further modify the fraction of the energy released on the star's surface.  A luminous spectral component emitted by  the  boundary layer\index{boundary 
layer} will exist in the X-ray spectrum of an accreting neutron star, in addition to the emission from the accretion disk. The shape of the spectra of luminous neutron  stars (soft state spectrum in Fig. \ref{fig:states_spe}) suggests that both accretion disk and boundary layer are in the optically thick regime.
As the luminosities of the two components are comparable, but the emitting area of the boundary layer is smaller than that of the accretion disk, its spectrum is expected to be  harder.

Looking from a different angle we may say that  because of the presence of the stellar surface, at the same $\dot{M}$ neutron stars are approximately twice more luminous than black holes. Indeed, in the case of the black hole accretion the kinetic energy of Keplerian motion at the inner edge of the accretion disk is nearly all advected into the black hole. 
If the central object is a neutron star, this energy is released in the boundary/spreading layer on its surface, approximately doubling the luminosity of the source.\footnote{A more accurate consideration should take into account geometry of the problem, namely the emission diagrams and orientation with respect to the observer of the emitting surfaces of the boundary layer\index{boundary layer} and accretion disk. This makes the boundary layer contribution to the observed emission  dependent  on  the inclination of the binary system, see for example \cite{ns_bl}.} An interesting consequence of this is that the ratio of $L_{Edd}/M$ is larger for neutron stars than for black holes.

\begin{figure}[htbp] 
\centerline{    
\hbox{
\includegraphics[width=0.51\hsize]{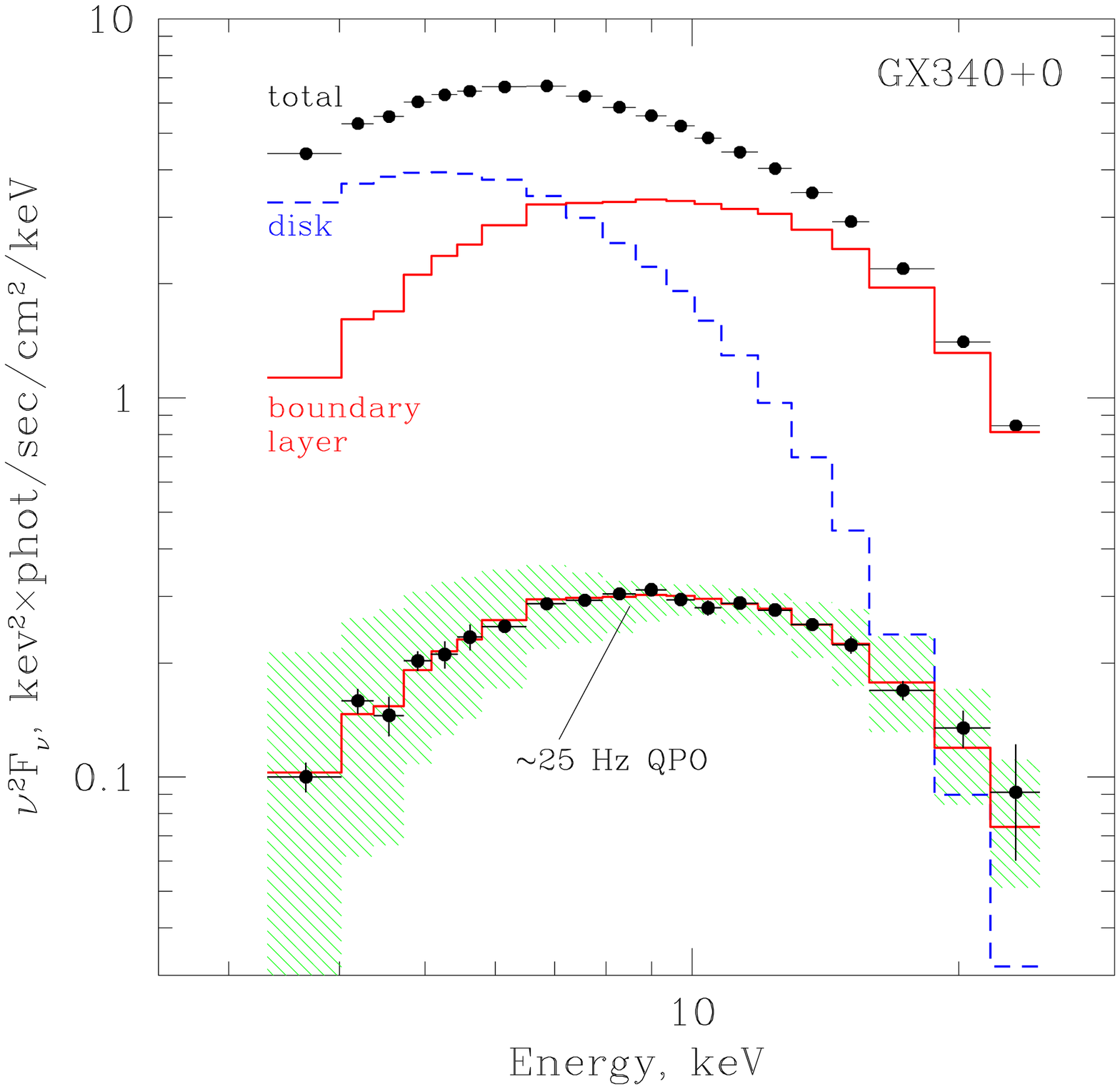}
\includegraphics[width=0.495\hsize]{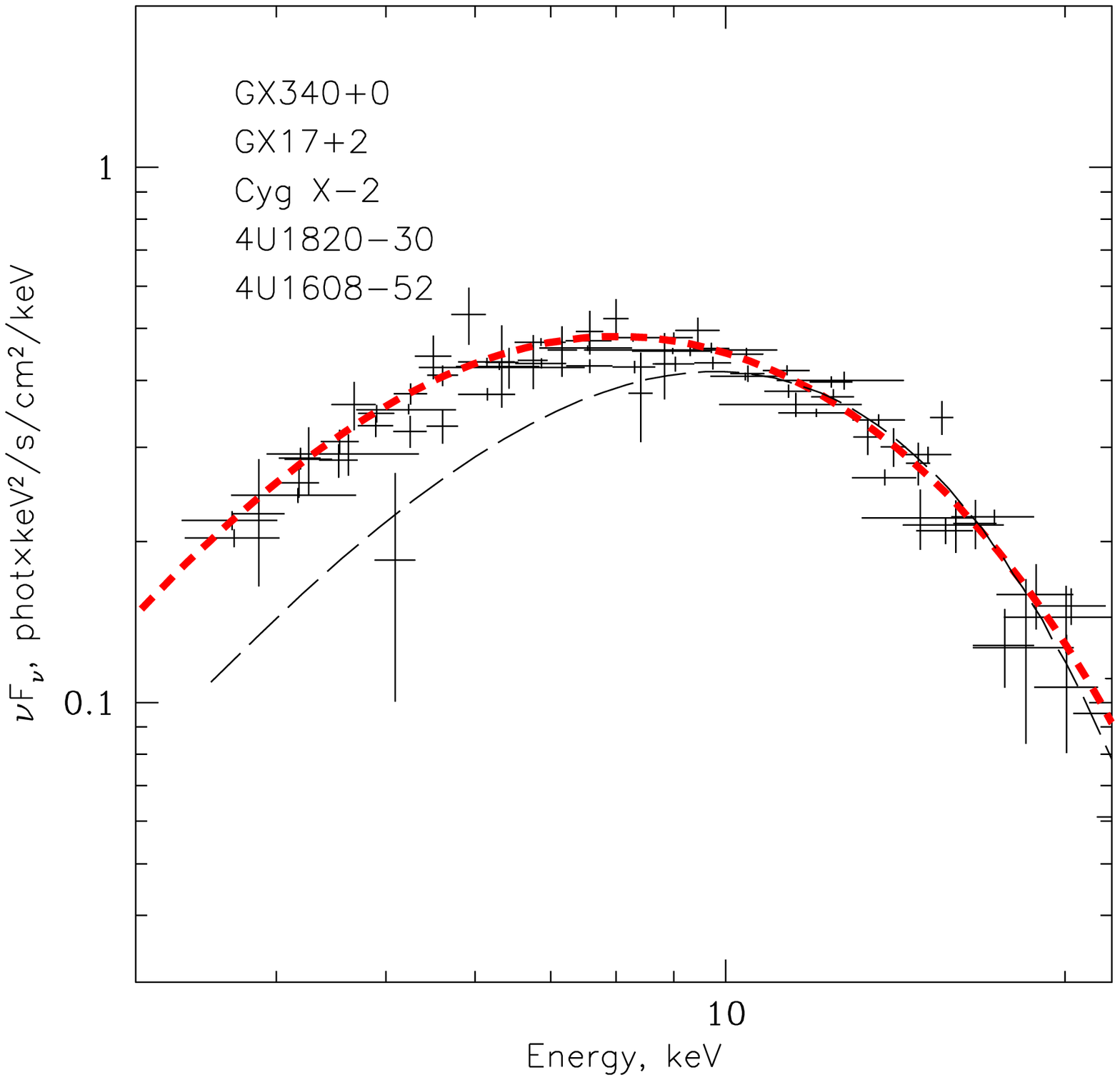}
}}
\caption{Left: Points with error bars showing total and frequency resolved-spectra of  GX~340+0. The dashed (blue in the color version of the plot) and upper solid (red) histograms show the  disk and boundary layer\index{boundary layer} spectra, the latter computed as a difference between the (observed) total and (predicted) accretion disk spectrum.  The lower solid histogram is obtained from the upper scaling it to the total energy flux of the frequency-resolved spectrum. Adopted from \cite{ns_bl}. 
Right: Fourier-frequency resolved spectra, corrected for the interstellar absorption ($\approx$boundary layer spectra) of 5 luminous accreting neutron stars.  For 4U~1608-52, the frequency-resolved spectrum of the lower kHz QPO\index{QPO} is shown. All spectra were corrected for the interstellar absorption. The thick short-dashed line shows the best fit Comptonization model with $kT_s=1.5$, $kT_e=3.3$ keV, $\tau=5$. The thin long-dashed line shows a blackbody spectrum with temperature $kT_{\rm bb}=2.4$ keV. Adopted from \cite{ns_bl2}.
}
\label{fig:bl}
\end{figure}

Due to the similarity of the spectra of the accretion disk and boundary layer\index{boundary layer}, the total spectrum has a smooth curved shape, which is difficult to decompose into separate spectral components.  This complicates analysis and interpretation of the neutron star spectra and, in spite of a very significant increase in the sensitivity of  X-ray instruments made in the last decade, still often leads to  ambiguous and contradicting results, even based on physically motivated spectral models. 
This degeneracy may be removed  with the help of timing information.
It has been noticed already in the early 80's that the variability patterns may be different for the boundary layer and accretion disk \cite{diskbb}.  Further progress has been made almost  twenty years later, thanks to the large collecting area of the PCA instrument aboard  RXTE and the use of novel data analysis techniques. Using the method of frequency-resolved spectral analysis it has been shown  that aperiodic and quasi-periodic variability of bright LMXBs -- atoll and Z- sources,  on $\sim$ sec -- msec time scales  is caused primarily by variations of the luminosity of the boundary layer\index{boundary layer} \cite{ns_bl}.
It was also shown that the boundary layer spectrum remains nearly constant in the course of  the luminosity variations  and is represented to certain accuracy by the  Fourier frequency resolved spectrum. This permits to resolve the degeneracy in the spectral fitting and  to separate contributions of the boundary layer and disk emission (Fig. \ref{fig:bl}, left panel). Interestingly, the spectrum of the boundary layer emission has the same shape in different objects and is nearly independent upon the global mass accretion rate in the investigated range of $\dot{M}\sim (0.1-1) \dot{M}_{\rm Edd}$ and in the limit of $\dot{M}\sim \dot{M}_{\rm Edd}$ is close to a Wien spectrum with $kT\sim 2.4$ keV (Fig. \ref{fig:bl}, right panel). Its independence on the global value of $\dot{M}$ lends support to the  theoretical suggestion by \cite{inogamov99} that the  boundary layer is radiation pressure supported. With this assumption, one can attempt to measure gravity on the neutron star surface and hence $M/R^2$, from the shape of the boundary layer spectrum, similarly to the photospheric expansion (i.e. Eddington limited) X-ray bursts. This gives results within the range of values obtained by other methods \cite{ns_bl2, suleimanov}.

As neutron stars have smaller mass than black holes,  the linear scale corresponding to the gravitational radius is smaller, $r_g\propto M$. This has two important consequences. Firstly, the surface area of the emitting region is $\propto M^2$, therefore at the same luminosity the temperature of black body emission will be larger, $T_{bb}\propto M^{-1/2}$. 
The spectrum is further modified by the contribution of the emission from the boundary layer\index{boundary layer}, which has yet smaller emitting region and harder spectrum. This  is illustrated by Fig. \ref{fig:states_spe} -- the soft state spectrum of the neutron star 4U~1705-44\index{4U~1705-44} is noticeably harder than that of the black hole Cyg~X-1\index{Cyg~X-1}.

Secondly, the smaller linear scale in neutron star accretion shifts the characteristic frequency scales by a factor $\propto M^{-1}$, suggesting $\sim 5$ times higher frequencies of variability in neutron star systems. This is confirmed by observations as demonstrated in the extensive comparison of  black hole and neutron star power density spectra  in \cite{nsbh_pds}. In addition, characteristic time scales in the boundary or spreading layer are significantly shorter than those in the accretion disk and are in the tens of kHz frequency domain. It has been suggested that a very high frequency component may exist in the power density spectra of neutron stars associated with the turbulence in the spreading layer \cite{nsbh_pds,inogamov99}. If detected, this may become a unique diagnostics tool of physical conditions in the spreading layer on the surface of the neutron star. So far, only upper limits have been obtained, they are at the level of $\sim 10^{-2}$ fractional rms \cite{nsbh_pds}.

The relatively cold surface of the neutron star is a source of copious soft photons. This is mostly relevant in the hard spectral state -- the low energy photons emitted by the  neutron star surface  result in a more efficient cooling of electrons in the Comptonization region, thus changing its energy balance and Comptonization parameter (luminosity enhancement factor, cf. Sect. \ref{sec:r-g}).  Consequently, the energy spectra of neutron stars in the hard state are significantly softer than those of black holes -- they have large spectral index (smaller Comptonization parameter)  and a smaller value of the high energy cut-off (lower electron temperature).  Both these effects are clearly seen in the spectra shown in  Figs. \ref{fig:states_spe},\ref{fig:bhns_spe}.

The existence of soft and hard spectral states in neutron stars suggests that, similarly to  black holes, the accretion flow can change its configuration from optically thin to optically thick. A remarkable fact is that this change seems to occur to both accretion disk and boundary layer\index{boundary layer} (quasi) simultaneously. Indeed, to my knowledge, no two-component (soft + hard) spectrum has been observed in the case of accreting neutron stars so far.

\end{document}